\documentclass[12pt, letterpaper]{article}
\usepackage{jheppub}

\usepackage{bibentry}

\usepackage[nostyling,legacy]{ajtex}
\newcommand{\bec}{\begin{cases}} 
\newcommand{\eec}{\end{cases}}

\title{Holographic Reconstruction of Bubbles}
 
\author[a]{Philipp Burda}
\emailAdd{philipp.burda@mail.huji.ac.il}

\author[b,c]{Ruth Gregory}
\emailAdd{r.a.w.gregory@durham.ac.uk}

\author[b]{and Akash Jain}
\emailAdd{akash.jain@durham.ac.uk}

\affiliation[a]{Racah Institute of Physics, Hebrew University, 
Jerusalem 91904, Israel}
\affiliation[b]{Centre for Particle Theory \& Department of Mathematical
Sciences, Durham University, South Road, Durham, DH1 3LE, UK}
\affiliation[c]{Perimeter Institute, 31 Caroline Street North, Waterloo, ON,
N2L 2Y5, Canada} 

\abstract{We discuss the holographic reconstruction of static thin bubble walls
in BTZ black hole geometries. We consider two reconstruction prescriptions
suggested in recent years: hole-ography and light-cone cuts, in the context of
thin bubble walls, and comment on their applicability in the presence of
non-trivial matter in the bulk. We find that while the light-cone cuts
prescription goes through within its own limitations, the current
hole-ographic approaches are inadequate to describe bubble spacetimes
completely. Much like entanglement shadows found around BTZ black holes and
conical defects in the bulk, we find that thin bubbles are accompanied by
shadows of their own, which are regions of spacetime which are only partially
probed by minimal geodesics. We speculate that such shadows might be a generic
feature of the presence of matter in the bulk.}

\preprint{DCPT-18/03}

\begin{document}
\maketitle 

\section{Introduction}\label{sec-intro}

Since the seminal work of Maldacena in 1997~\cite{Maldacena:1997re}, holography
has came to become one of the cornerstones of twenty-first century high energy
physics. The so called ``AdS/CFT duality'' has found a multitude of applications in
condensed matter physics, as it provides a probe for studying strongly coupled
highly quantum phenomenon in field theories using their dual weakly coupled
semi-classical gravity theories. These applications have been well explored in
the literature; see e.g.~\cite{Hartnoll:2009sz, McGreevy:2009xe, Gubser:2009md}
and references therein. The other direction of this duality can in turn be used
to potentially understand some characteristics of quantum gravity --- which
continues to be one of the most profound puzzles in physics --- by looking at
the well understood weakly coupled field theories. Soon after the AdS/CFT conjecture
was proposed, these ideas began to take shape under the name of bulk
reconstruction~\cite{Balasubramanian:1999ri,Giddings:1999qu}. Since then, there
has been a tremendous amount of research towards reconstructing local bulk
operators from the boundary CFT ones~\cite{Hamilton:2006az, Hamilton:2005ju,
Hamilton:2006fh, Hamilton:2007wj, Almheiri:2014lwa, Swingle:2012wq,
Qi:2013caa}. See~\cite{DeJonckheere:2017qkk} for an excellent set of lecture
notes on the subject, and references therein.

An important part of the bulk reconstruction program is to understand how
classical spacetime emerges from the underlying quantum degrees of
freedom. Presumably, any theory of quantum gravity should provide an answer to
this question, however, the precise mechanism is still unknown. In the
holographic setting, one would like to isolate the degrees of freedom in the
boundary field theory that might encode the information about the geometry, or
more specifically the metric, of the bulk spacetime.  A number of proposals have
been put forth in this regard over the past two decades, perhaps the most
developed of which is the idea that the bulk spacetime emerges from the
entanglement structure of the boundary field theory~\cite{Ryu:2006bv,
VanRaamsdonk:2009ar}. In $(2+1)$ bulk dimensions, which will be the focus of
this paper, these ideas have been rigorously developed using boundary
observables like entanglement entropy, differential entropy and
entwinement~\cite{Balasubramanian:2013lsa, Czech:2014ppa, Myers:2014jia,
Headrick:2014eia, Balasubramanian:2014sra, Balasubramanian:2016xho}. Another
more recent approach to bulk reconstruction, called light-cone cuts, uses $n$-point
correlation functions in the boundary field theory to obtain the bulk metric in
arbitrary number of dimensions up to a conformal
factor~\cite{Engelhardt:2016wgb}.

Most of the work cited above for geometric bulk reconstruction trials the
methods with a fairly limited class of bulk geometries: global AdS$_{3}$, 
BTZ black holes and
point conical defects, all of which are quotients of AdS$_{3}$. It is important
therefore to subject these proposals to further tests, and in particular, to
apply these recipes to more complicated (and hence realistic) geometries, which
will allow us to understand them better and help determine their limits of
applicability. To this end, in this paper we consider spacetimes with massive
thin shells as an interesting candidate to test the validity of the proposed
bulk reconstruction approaches. These spacetimes have several generic features
related to the problem: the presence of non-trivial matter in the bulk in the
form of a collapsing or expanding thin shell is combined with \emph{almost} the
same amount of isometries as in the empty AdS spacetime. However crucially, the
presence of the shell-matter in the bulk causes a controllable breaking of the
bulk symmetries that can lead to interesting features in the dual boundary field
theory.

There are two type of physical processes which could be described by massive
thin shells. A collapsing shell describes the process of a spherical collapse
and black hole formation in AdS spacetime. This would correspond to a
thermalisation process in the boundary field theory and has been explored e.g.\
in~\cite{Danielsson:1999zt, Danielsson:1999fa}. Thin wall bubbles also 
describe the process of false vacuum decay
\cite{Coleman:1977py,Coleman:1980aw},
can be time dependent or static, and can occur with or 
without a black hole \cite{Gregory:2013hja,Burda:2015isa,Burda:2015yfa}.  
Aspects of their dual holographic description were studied
in~\cite{Barbon:2010gn}.  We will concentrate on the special case at the
boundary of these two examples --- static shells in $(2+1)$-dimensions. These can
be viewed as special instantons in the context of vacuum decay, or as limits of
flows or domain walls in AdS.  The advantage of using a thin wall is that it is
an analytic gravitational set-up. We can therefore derive analytic expressions
for the quantities being proposed in bulk reconstruction, and easily explore
their validity.

Our set-up is such that the static shell solution bounds two BTZ black hole
spacetimes: the \emph{inside} ``$-$'' and \emph{outside} ``$+$''
respectively. These two spacetimes in general have different mass parameters,
$M_{-} \neq M_{+}$, and different AdS radii,
$\ell_{-} < \ell_{+}$.\footnote{This last inequality between AdS radii follows
from the requirement that $\Lambda_{-} < \Lambda_{+}$ for tunnelling towards a
deeper vacuum. In the boundary field theory it corresponds to the proper
direction of an RG flow~\cite{Freedman:1999gp}.} We refer to these spacetimes
as BTZ bubbles, and it is clear that they cover a wide range of geometries,
easily generalisable to higher dimensions. Leaving consideration of this, and
dynamical shells, for the future, it is worth mentioning that time evolution of
the holographic entanglement entropy in higher dimensions with collapsing shells
has previously been studied in the adiabatic limit
in~\cite{Keranen:2015fqa}. However, the question of bulk reconstruction was not
addressed there.

We will focus on two proposals for bulk reconstruction in this paper that use a
geometrical approach to the problem, in the sense that they define points and
distances in the bulk via introduction of auxiliary geometric constructions on
top of the boundary field theory data. The first approach was introduced
in~\cite{Balasubramanian:2013lsa, Czech:2014ppa}, which we will refer to as
``hole-ography''. This method provides a recipe to reconstruct the spatial part
of the bulk metric using the entanglement structure of the boundary CFT, which
involves observables like entanglement entropy, differential entropy and
entwinement. This proposal is quite natural, as the very first indication of an
``emergence'' of the bulk geometry was seen in the Ryu-Takayanagi formula for
holographic entanglement entropy~\cite{Ryu:2006bv}. It has been widely felt that
the entanglement entropy data of the boundary field theory should play an
important role in bulk reconstruction, therefore, we discuss holographic
entanglement entropy for the BTZ bubble spacetimes in details in
\cref{sec-HEE}. The hole-ography method itself is reviewed and applied to BTZ
bubble spacetimes in \cref{holeography}.

The second and more recent approach of using \emph{light-cone cuts} to
reconstruct bulk geometry was introduced in~\cite{Engelhardt:2016crc,
Engelhardt:2016wgb}. This method provides a strategy to obtain the metric of
the bulk spacetime up to an overall conformal factor. However, only the part of
the spacetime that is in a casual contact with the boundary can be reconstructed
using this prescription. The method is based on the knowledge of the divergence
structure of the correlation functions in the boundary CFT.\ We will review this
method in details in \cref{LightConeCuts} and provide explicit examples of the
reconstruction of empty AdS, BTZ black hole and BTZ bubble spacetimes.

In addition to the comments on reconstructability of bubbles, we note the
existence of what we call ``bubble shadows''. These are a region of spacetime
surrounding a bubble in the bulk, which can be seen as a particular
generalisation of entanglement shadows discussed extensively in the literature
(see e.g.~~\cite{Balasubramanian:2014sra, Freivogel:2014lja,
  Balasubramanian:2017hgy}). Unlike an entanglement shadow however, which is a
region of spacetime where minimal length geodesics (Ryu-Takayanagi surfaces) do
not enter, these bubble shadows are only partially probed by minimal
geodesics. Given that entanglement shadows have been found in BTZ black holes
and spacetimes with a conical defect, which can be seen as point matter sources
in the bulk, it seems to suggest that such shadows might be a generic feature of
the presence of matter in the bulk.

The paper is structured as follows. We give an overview of static thin wall
bubble geometries and their geodesics in \cref{sec:staticbubbles}. In
\cref{sec-HEE} we give a detailed discussion of holographic entanglement entropy
in the presence of bubbles in the bulk, and note the existence of bubble
shadows. In \cref{sec:bubble-reconstruction} we investigate hole-ographic and
light-cone cuts reconstruction schemes in the context of bubble
spacetimes. Finally, we close with discussion in \cref{sec:discussion}. In the
appendix, we present the construction of kinematic spaces associated with bubble
spacetimes.


\section{Static bubble geometry}\label{sec:staticbubbles}

We will start with a brief review of thin wall bubbles and Israel junction
conditions (see~\cite{Gregory:2008rf} for more details). We will work out
various kinds of geodesics in this geometry, which will form the basis for our
discussion in the bulk of this work. A generic bubble spacetime in arbitrary
dimensions consists of an infinitesimally thin wall separating two bulk
solutions of the vacuum Einstein's equations. The symmetries of our setup and
bubble wall energy-momentum require that each bulk has the form of an AdS black
hole~\cite{Bowcock:2000cq}. The vacuum energy and mass parameters of the black
hole on each side are, in general, different and related to the tension of the
bubble wall via the Israel junction
conditions~\cite{Israel:1966rt}. Generically, a bubble will have a
time-dependent trajectory, however, to illustrate the issues in geometric bulk
reconstruction, it is sufficient to consider the subset of bubble geometries
that are static. We therefore briefly review the $(2+1)$-dimensional bubble
geometries in this section before considering their holographic interpretation.

\subsection{Static bubble metric}

Recall that the metric of a $(2+1)$-dimensional BTZ black hole can be written as
\begin{equation}\label{the-metric}
  ds^2 = \frac{1}{z^2}\lb - \lb 1 - M z^2 \rb dt^2 + dx^2 +
  \frac{\ell^2}{1 - M z^2} dz^2 \rb,
\end{equation}
with coordinates $t,x\in \bbR$ and $z \in(0, 1/\sqrt{M})$. Here $\ell$ is the
AdS radius related to the cosmological constant as $\Lambda = -1/\ell^{2}$ and
$M$ is the mass parameter of the black hole. Notice that this metric is
invariant under the scaling of coordinates $\{t,x,z\} \to \{"l t,"l x,"l z\}$ if
we transform $M \ra M /"l^{2}$. The horizon of the black hole is at
$z = 1/\sqrt{M}$, while $z \to 0$ represents the asymptotic boundary of the
spacetime.

It is worth pointing out that our coordinate $x\in \bbR$, and hence the horizon
of the black hole, is non-compact. This is in contrast with most of the
holographic literature on this subject (see e.g.~\cite{Ryu:2006bv, Ryu:2006ef}),
where $x\in \rmS^1$ is taken to be compact. An important feature of these
compact black holes, as is well illustrated in~\cite{Ryu:2006bv, Ryu:2006ef}, is
that for any interval at the boundary, there is an infinite cascade of geodesics
anchored at its end points, characterised by their winding number around the
black hole. For a major part of this work, we will be interested in holographic
entanglement entropy, wherein the length of the shortest among these infinite
set of geodesics computes the entanglement entropy of the boundary interval in
question, while the longer ones are said to compute
``entwinement''~\cite{Balasubramanian:2014sra}. As we shall explore in the due
course, the presence of a bubble gives rise to two new geodesics for some
boundary intervals, on top of the infinite cascade arising due to
compactness. Therefore by going to a non-compact version of the BTZ black hole,
which is essentially an infinite cover of the compact one, we can efficiently
isolate and focus on the effects of the bubble.

To construct a bubble spacetime, we will ``glue'' together two BTZ black holes
with masses $M_\pm$ and AdS radii $\ell_\pm$ along a timelike hypersurface given
by $z=Z(\tau)$, while respecting the translation invariance in $x$
direction. Here $\tau$ is a timelike coordinate on the hypersurface. The
geometry is supported by a brane/bubble with constant tension $\s$ on the
hypersurface. By a suitable choice of coordinates, we take an ansatz for the
metric
\begin{equation}\label{bubble.metric}
ds^2 = 
\begin{cases}
\dsp \frac{1}{ z^2}\lb - (1-M_+ z^2) dt^2_+ + dx^2
+ \frac{\ell_+^2}{1-M_+ z^2} dz^2 \rb
&\quad \text{for} \quad  z \leq Z(\t), \vspace{0.1cm} \\
\dsp \frac{1}{ z^2}\lb - (1-M_- z^2) dt^2_- + dx^2
+ \frac{\ell_-^2}{1-M_- z^2} dz^2 \rb
&\quad \text{for} \quad  z \geq Z(\t).
\end{cases}
\end{equation}
The bulk time coordinates are given by $t_{\pm}$, which becomes a function of
$"t$ in the vicinity of the bubble.  For this ansatz to be consistent, it should
induce the same metric on the bubble $z=Z(\tau)$ from either sides. If we
represent the induced metric on the bubble as
\begin{equation}
  ds^2_{\text{Bubble}} = h_{ij} dx^i dx^j = \frac{1}{{Z(\t)}^2} \lb - d\t^2 + dx^2 \rb,
\end{equation}
this gives a consistency condition on the bulk metric ansatz
\begin{equation}\label{propertime}
(1-M_\pm Z^2) \dot t^2_{\pm} - \frac{\ell_\pm^2 \dot Z^2}{(1-M_\pm Z^2)} = 1,
\end{equation}
where dot denotes the derivative with respect to $\t$. This condition should be
read as relating the time coordinates $t_\pm$ to $"t$. In particular for a
static bubble, defined by $\dot Z = \ddot Z = 0$, this boils down to a simple relation
\begin{equation}
t_+ \sqrt{1 - M_+ Z^2} = t_-\sqrt{1 - M_- Z^2} = "t.
\end{equation}
Except on the bubble, metric in \cref{bubble.metric} is merely a BTZ black hole
and hence satisfies the Einstein equations. On the bubble however, Einstein
equations imply that this geometry can be supported by a uniform tension bubble
with an energy-momentum tensor $T^{i}_{\ j} = -"s "d^i_{\ j}$, provided the
Israel junction conditions~\cite{Israel:1966rt} are met
\begin{align}\label{E-P}
8\pi G \s &= - \sqrt{\frac{1-M_+Z^2}{\ell_+^2} + \dot Z^2}
+ \sqrt{\frac{1-M_- Z^2}{\ell_-^2} + \dot Z^2}, \nn\\
\frac{\ddot Z - \frac{M_+ Z}{\ell_+^2}}{\sqrt{\frac{1-M_+Z^2}{\ell_+^2} + \dot Z^2}}
&= \frac{\ddot Z - \frac{M_- Z}{\ell_-^2}} {\sqrt{\frac{1-M_-Z^2}{\ell_-^2} + \dot Z^2}}.
\end{align}
For the static case these conditions simplify considerably, and imply a range of
parameter space where we are allowed to have a static bubble. For
$M_\pm \neq 0$, \cref{E-P} in the static case implies
\begin{equation}\label{E-P_static}
8\pi G \s = \sqrt{\lb \frac{M_-}{\ell_-^2} - \frac{M_+}{\ell_+^2} \rb 
\lb \frac{1}{M_-} - \frac{1}{M_+} \rb }, \qquad
Z = \frac{1}{\sqrt{M_+M_-}}\sqrt{\frac{\frac{M_-^2}{\ell_-^2} 
- \frac{M_+^2}{\ell_+^2}}{\frac{M_-}{\ell_-^2} - \frac{M_+}{\ell_+^2}}}.
\end{equation}
Requiring the bubble tension $\s\geq 0$ and radius
$0 < Z < \min(\frac{1}{\sqrt{M_+}}, \frac{1}{\sqrt{M_-}})$, it gives the allowed
range of parameter space as
\begin{equation}
M_+ \geq M_-, \qquad
\frac{\ell_+}{M_+} \geq \frac{\ell_-}{M_-},
\label{mpmconstraints}
\end{equation}
which also implies a weaker condition $\ell_+ \geq \ell_-$. On the other hand,
if either of the black hole masses $M_+$ or $M_-$ is zero, the static condition
forces the other mass to vanish as well. Consequently for $M_\pm = 0$, using
\cref{E-P} we get
\begin{equation}
8\pi G \s = \frac{1}{\ell_-} - \frac{1}{\ell_+}, \qquad Z \in \bbR^{+},
\end{equation}
with the allowed region of parameter space
\begin{equation}
\ell_+ \geq \ell_-.
\end{equation}
It is interesting to note that in both the cases, the bubble separates a BTZ
spacetime with a less negative cosmological constant and higher mass parameter
in the UV (near the boundary) from a more negative cosmological constant and
lower mass parameter in the IR (deep in the bulk). This is in agreement with a
holographic c-theorem~\cite{Freedman:1999gp, Myers:2010tj}.

\subsection{Spatial geodesics}
\label{bubble-geodesics}\label{spatial-geodesics} 

During our discussion of bulk reconstruction later, we will extensively need the
form of geodesics in the bulk. Hence we dedicate this subsection and the next to
derive geodesics in static bubble geometries. The bubble spacetime is locally a BTZ
black hole everywhere except in the vicinity of the bubble. So to find the
geodesics we can use the following trick: we can start with the known geodesics
in the ``$+$'' and ``$-$'' parts of the spacetime independently, and ``glue''
them with suitable boundary conditions (corresponding to the local continuity
and smoothness of the geodesic). As innocuous as it sounds, this procedure can
be quite cumbersome for a generic geodesic. Fortunately, for our purposes it
suffices to consider just two special cases: spatial geodesics confined to a
constant time slice and null geodesics that reach out to the boundary.

By a straightforward computation, one finds that spatial geodesics on a constant
time slice of the BTZ metric in \cref{the-metric} are given by two distinct
branches. First kind of geodesics start from the boundary, turn at a point
$(t_0,x_0,z_0)$ in the bulk and return to the boundary:
\begin{equation}
t = t_0, \qquad
x = x_0 \pm
\frac{\ell}{\sqrt M} \sinh^{-1} \lb\sqrt{\frac{M(z_0^2-z^2)}{1-Mz_0^2}} \rb
\quad \text{with} \quad 0 < z_0 < \frac{1}{\sqrt M}.
\end{equation}
These are the geodesics one would consider when computing entanglement entropy
of a spatial slice at the boundary. The other kind of geodesics start from the
boundary, cross the black hole horizon and escape all the way to the other
asymptotic boundary:
\begin{equation}
t = t_0, \qquad
x = x_0 \pm
\frac{\ell}{\sqrt M} \cosh^{-1} \lb\sqrt{\frac{M(z_0^2-z^2)}{Mz_0^2-1}} \rb
\quad \text{with} \quad \frac{1}{\sqrt M} < z_0 < \infty.
\end{equation}
One would employ these if one needs to understand entanglement between the two
asymptotic regions. If we are interested in the intervals at a boundary of the
BTZ black hole, they are not quite as useful. However, upon introduction of
bubbles, we find that they in fact start to play an important role.

\begin{figure}[t]
\centering
\begin{subfigure}[b]{0.49\textwidth}
\includegraphics[width=\textwidth]{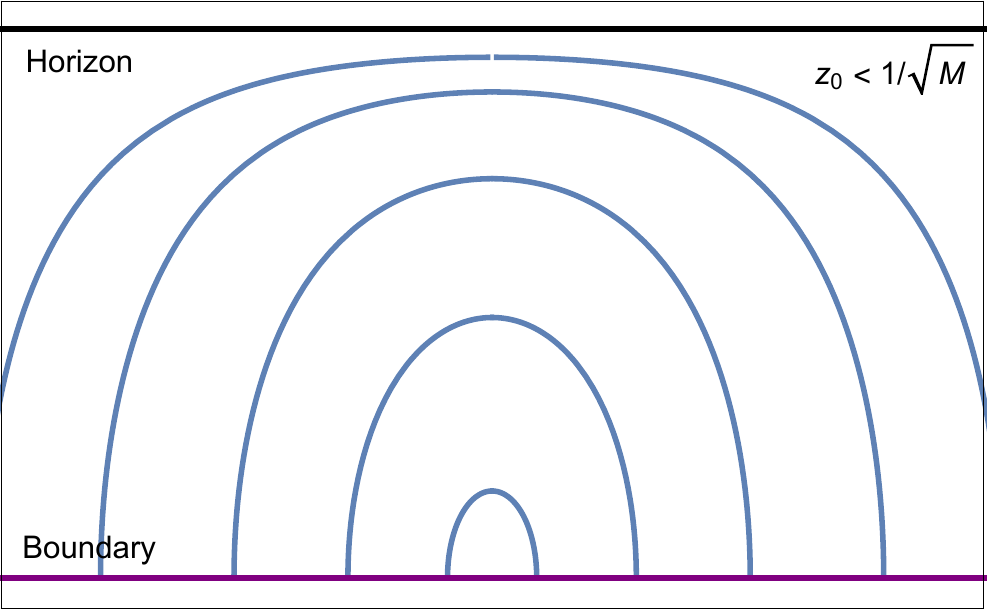}
\caption{}
\end{subfigure}
\begin{subfigure}[b]{0.49\textwidth}
\includegraphics[width=\textwidth]{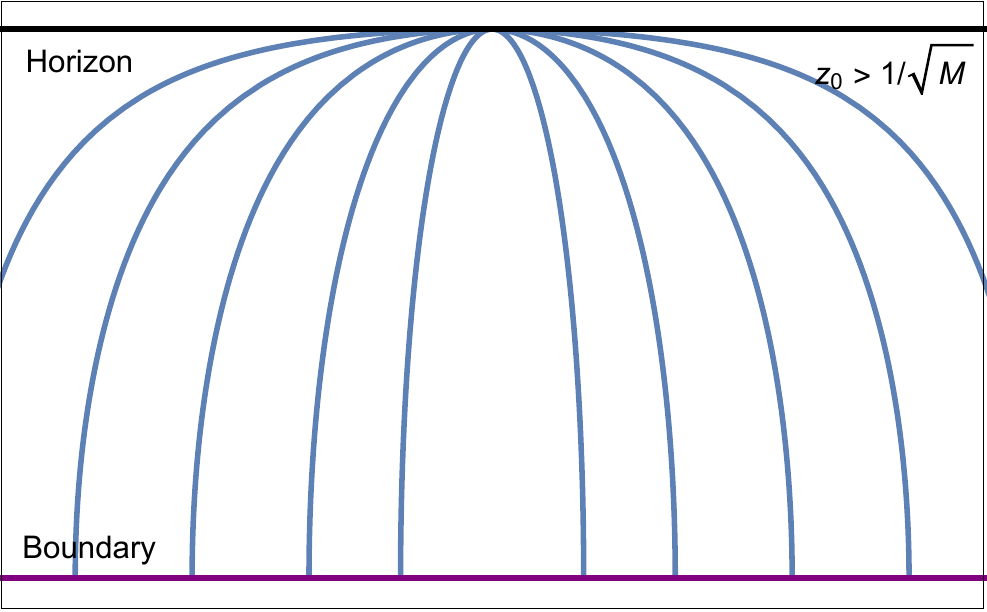}
\caption{}
\end{subfigure}
\caption{\label{BTZ_geodesics} Spatial geodesics for BTZ black holes with
\textbf{(a)} $0 < z_0 < 1/\sqrt M$ and \textbf{(b)}
$1/\sqrt M < z_0 < \infty$. [Parameters: $\ell = 1$, $M = 0.1$, $t_0 = 0$
and $x_0 = 0$]. Note that the second branch of geodesics appears to be
grazing the horizon and returning, but this is just a product of an
inappropriate set of coordinates. They indeed cross the horizon, as can be
illustrated by computing their slope at the horizon.}
\end{figure}

For notational clarity, let us combine the two branches of geodesics 
into an analytically continued
form
\begin{equation}\label{geodesics.BTZ}
t = t_0, \qquad
x = x_0 \pm \frac{\ell}{\sqrt M} "cS^{-1} \lB z,z_0,M \rB \,,
\end{equation}
where
\begin{equation}\label{calSdefn}
{\cal {S}}^{-1} \left [ z,z_0,M \right ]
= \sinh^{-1}\lb \sqrt{\frac{M( z_0^2- z^2)}{1-Mz_0^2}} \rb
- \frac{i"p}{2} "Q({\textstyle |z_0| - \frac{1}{\sqrt M}}).
\end{equation}
Here $0 < z_0 < \infty$ and $"Q(z)$ is the Heaviside theta function. We have
illustrated some representative geodesics in \cref{BTZ_geodesics}.

We construct spatial geodesics in the static bubble geometry by gluing spatial
geodesics in the ``$+$'' and ``$-$'' spacetimes, given in \cref{geodesics.BTZ},
and requiring continuity and smoothness of the geodesic on the bubble. If the
geodesic turns at a point $(t_0, x_0, z_0)$ in ``$+$'' spacetime with
$ z_0 \leq Z$, it does not enter the ``$-$'' spacetime at all, and is simply
given by
\begin{equation}\label{out-geodesic}
t_+ = t_{0}, \qquad
x = x_{0} \pm \frac{\ell_+}{\sqrt{M_+}} {\cal{S}}^{-1}\left [ z, z_0, M_+ \right ] \,. 
\end{equation}
On the other hand if $z_0>Z$, then the geodesic will enter the minus spacetime,
thus crossing the bubble at two distinct points. Although the spatial
coordinates $x$ and $z$ have been chosen to be continuous across the bubble
wall, the time coordinates are different on each side, so the time $t_+=t_0$ on
the plus side of the bubble will match up to $t_- = "g t_0$ on the minus side,
where $"g = \sqrt\frac{1-M_+ Z^2}{1-M_- Z^2}$ is the time warp factor when we
cross the wall. Once on the minus side of the bubble, the geodesic will have the
standard form
\begin{equation}
t_- = "g t_{0}\;, \qquad
x = x_{0} \pm  \frac{\ell_-}{\sqrt{M_-}} "cS^{-1} \lB z,z_0,M_-\rB,
\end{equation}
that must be matched through the bubble to a generic geodesic segment in the
plus spacetime
\begin{equation}
t_+ = t_{0} \pm \a_1\;,\qquad
x = x_{0} \pm \a_2 \pm \frac{\ell_+}{\sqrt{M_+}} {\cal{S}}^{-1} \lB z,\alpha_3,M_+\rB.
\end{equation} 
The constants $\a_{1}$, $\a_2$ and $\a_3$ are fixed by local continuity and
smoothness of the geodesic at the bubble
\begin{equation}
\a_1 = 0, \qquad
\a_2 = \frac{\ell_-}{\sqrt M_-} {\cal{S}}^{-1} \lB Z,z_0,M_- \rB-
\frac{\ell_+}{\sqrt M_+} {\cal{S}}^{-1} \lB Z,z_0,M_+ \rB , 
\qquad \a_3 =  z_0.
\end{equation}
These geodesics start from the boundary in ``$+$'' spacetime and reach the
bubble, then (in this coordinate system) ``refract'' through the bubble to a
standard ``$-$'' geodesic. Depending on how $z_0$ compares to $1/\sqrt{M_-}$,
these geodesics will either turn and return via a similar trajectory, or will
cross the horizon all the way to the other asymptotic boundary.

Note that the geodesics in bubble geometry only cross the horizon if
$z_0 > 1/\sqrt{M_-}$. In particular if $M_+ > M_-$, geodesics with
$1/\sqrt{M_+} < z_0 < 1/\sqrt{M_-}$ will stay well away from the black
hole. These geodesics have segments in the ``$+$'' spacetime which would have
crossed the horizon by themselves in the absence of the bubble, but the bubble
refracts them such that they do not make it to the horizon after all.

\begin{figure}[t] \centering
\begin{subfigure}[b]{0.49\textwidth}
\includegraphics[width=\textwidth]{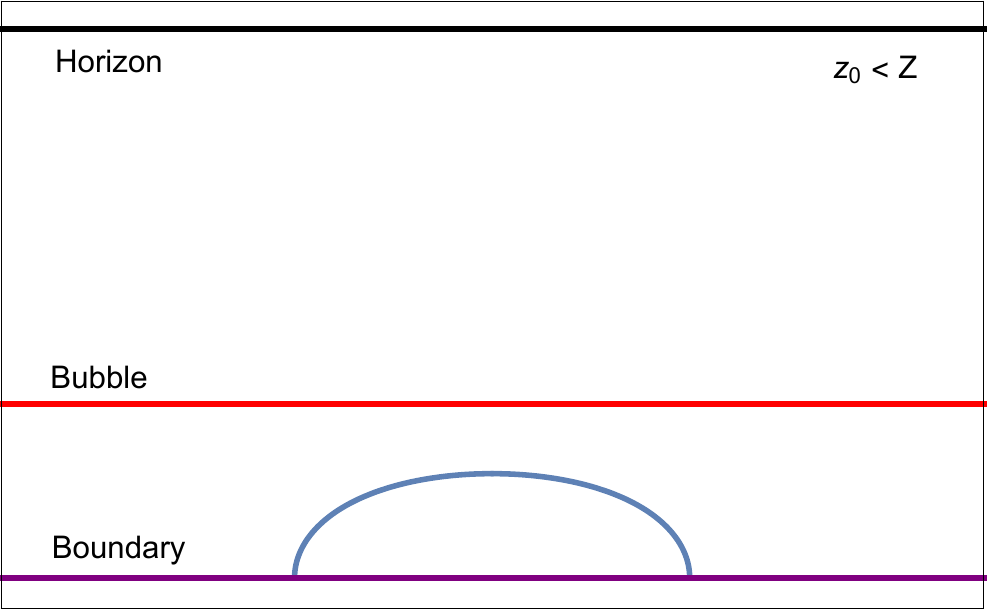}
\caption{}
\end{subfigure}
\begin{subfigure}[b]{0.49\textwidth}
\includegraphics[width=\textwidth]{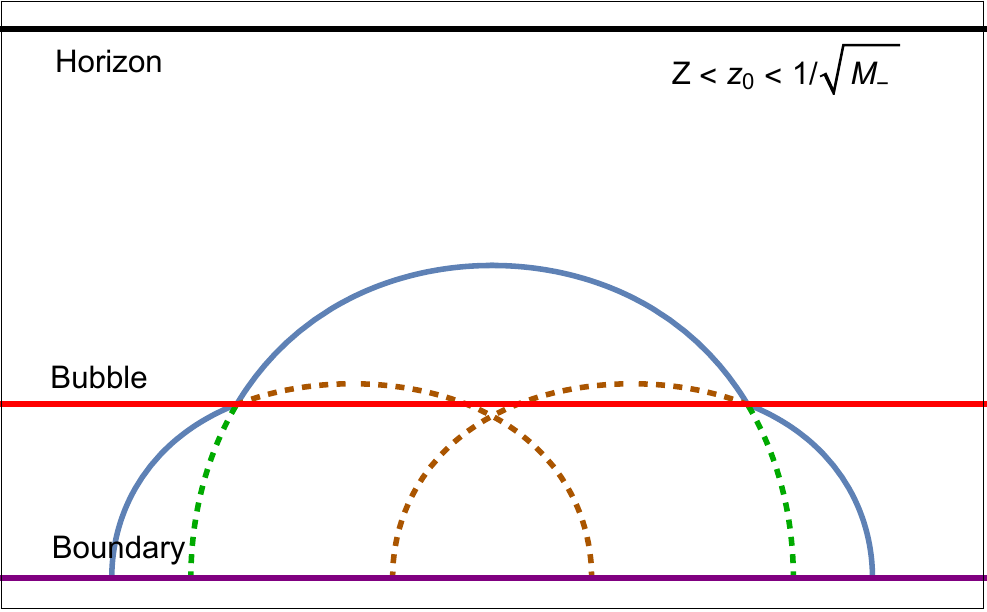}
\caption{}
\end{subfigure}
\caption{\label{fig.BTZgeodesics} Sample spatial geodesics for BTZ bubbles for
\textbf{(a)} $ z_0 < Z$ and \textbf{(b)} $Z< z_0<1/\sqrt{M_-}$. Dotted lines
represent geodesics in ``$+$'' and ``$-$'' spacetimes, which have been
``glued'' together to get the resultant geodesics in bubble
spacetime. [Parameters: $\ell_+ = 2$, $M_+ = 0.8$, $\ell_- = 1$,
$M_- = 0.1$, $Z = 1$, $t_0 = 0$, $x_0 = 0$ and $ z_0 = 0.6,~1.8$].
}
\end{figure}

To summarize, spatial geodesics in bubble geometry are given in the ``$+$''
spacetime as
\begin{equation}\label{cross-geodesic-2}
\begin{aligned}
t_+ &= t_{0},\\
x &= x_{0} \pm \frac{\ell_+}{\sqrt{M_+}} {\cal{S}}^{-1} \lB z,z_0,M_+ \rB\\
& \qquad
\pm \Theta({\textstyle z_0 - Z}) \lb \frac{\ell_-}{\sqrt{M_-}} {\cal{S}}^{-1} \lB Z,z_0,M_- \rB
- \frac{\ell_+}{\sqrt{M_+}} {\cal{S}}^{-1} \lB Z, z_0, M_+ \rB \rb,
\end{aligned}
\end{equation}
while for $z_0 > Z$ they also have a branch in the ``$-$'' spacetime
\begin{equation}\label{cross-geodesic-1}
t_- = "g t_{0},\qquad
x = x_{0} \pm  \frac{\ell_-}{\sqrt{M_-}} {\cal{S}}^{-1} \lB z,z_0,M_- \rB.
\end{equation}
These different types of spatial bubble geodesics are shown in
\cref{fig.BTZgeodesics}. As discussed in~\cite{Balasubramanian:2011ur}, the
physical effect of the bubble and the interior BTZ spacetime is analogous to a
medium with lower refractive index to the exterior BTZ geometry.  Geodesics
therefore have a tendency to cross to the interior ``$-$'' spacetime to transit
across the bulk. This gives rise to interesting phenomena when considering the
length of such geodesics, which we will review in \cref{sec-HEE} while talking
about holographic entanglement entropy.

\subsection{Null geodesics}\label{null-geodesics}

We now move on to the discussion of null geodesics. A generic null geodesic for
the BTZ black hole metric (\ref{the-metric}), which escapes to the boundary, is
given as,
\begin{equation}\label{nullgeod.BTZ}
\begin{split}
t &= t_0 \pm \frac{\ell}{\sqrt M} "cT_t^{-1}[z,p,M], \quad
"cT_t^{-1}[z,p,M] = \tanh^{-1} \lb \frac{ z \sqrt M}{\sqrt{1- p^2 (1-Mz^2)}} \rb, \\
x &= x_0 \pm \frac{\ell}{\sqrt M} "cT_x^{-1}[z,p,M], \quad
"cT_x^{-1}[z,p,M] =  \tanh^{-1} \lb \frac{p z \sqrt M}{\sqrt{1- p^2 (1-M z^2)}} \rb.
\end{split}
\end{equation}
Here $(t_0,x_0)$ are coordinates of a point at the boundary where the geodesic
hits, while $p\in(-1,1)$ represents the momentum of the null geodesic in $x$
direction. We can also find null geodesics which do not escape to the boundary,
but we will not need them in this work.

Now performing an analysis similar to that for the spatial geodesics,
i.e. considering null geodesics in ``$+$'' and ``$-$'' spacetimes and imposing
suitable boundary conditions on the bubble, we can recover null geodesics for
static BTZ bubbles. The resultant solutions for the ones that escape to the
boundary are: for $z < Z$ in ``$+$'' spacetime
\begin{equation}\label{nullgeod.Bubble.outside}
\begin{split}
t_+ &= t_0 \pm \frac{\ell_+}{\sqrt{M_+}} \Big( "cT_t^{-1}[z,p_0,M_+]
- "cT_t^{-1}[Z,p_0,M_+] \Big)
\pm \frac{1}{"g} \frac{\ell_-}{\sqrt{M_-}} "cT_t^{-1}[Z,"g p_0,M_-], \\
x &= x_0 \pm \frac{\ell_+}{\sqrt{M_+}} \Big( "cT_x^{-1}[z,p_0,M_+]
- "cT_x^{-1}[Z,p_0,M_+] \Big)
\pm \frac{\ell_-}{\sqrt{M_-}} "cT_x^{-1}[Z,"g p_0,M_-],
\end{split}
\end{equation}
and for $z\geq Z$ in  ``$-$'' spacetime
\begin{equation}\label{nullgeod.Bubble.inside}
t_- = "g t_0
\pm \frac{\ell_-}{\sqrt M_-} "cT_t^{-1}[z,"g p_0,M_-], \qquad
x = x_0 \pm \frac{\ell_-}{\sqrt{M_-}} "cT_x^{-1}[z,"g p_0,M_-].
\end{equation}
Here $(t_0,x_0)$ are coordinates of a point at the boundary where the geodesic
hits, while $p_0$ and $"g p_0$ represent the momentum of the null geodesic in
the $x$ direction in ``$+$'' and ``$-$'' spacetime respectively.

This finishes our general discussion of static bubble geometries. We constructed
static uniform tension infinitesimally thin bubbles with a BTZ black hole
geometry on either sides, and studied the behavior of some special geodesics in
this spacetime. In the following sections we will use these results to probe
some exciting holographic implications of these bubbles.

\section{Holographic entanglement entropy}\label{sec-HEE}

In recent years, there has been a tremendous amount of interest in connections
between quantum gravity and quantum information theory. We have learned that we
can get some profound insights into the quantum nature of gravity by appealing
to techniques pertaining to quantum information~\cite{Ryu:2006bv,
Lewkowycz:2013nqa, Maldacena:2013xja, Brown:2015bva}. Perhaps the best
understood of these insights come from a field theory observable called the
``entanglement entropy''. Naively, entanglement entropy $S_{A}$ of a spatial
region $A$ in a field theory is a measure of quantum entanglement between
degrees of freedom living in $A$ and those in its complement. For an excellent
review on the subject, see~\cite{Rangamani:2016dms}. For field theories
which admit a holographic dual, entanglement entropy can be computed using the
formula due to Ryu-Takayanagi~\cite{Ryu:2006bv,Hubeny:2007xt}
\begin{equation}
S_A = \min_i\lb\frac{\text{Area}(\Sigma^i_A)}{4G} \rb.
\end{equation}
Here $\Sigma^i_A$ are extremal area surfaces in the bulk anchored at $A$ at the
boundary, i.e.\ $\dow\Sigma_A^i = \dow A$, and are homologous to $A$ (can be
smoothly deformed into $A$ at the boundary). The index ``$i$'' runs over
multiple such surfaces, if available, in which case the formula picks up the
minimal area surface. For a $(1+1)$-dimensional field theory which has a
$(2+1)$-dimensional bulk dual, like the ones we are interested in, spatial
region $A$ is just an interval at the boundary and $\Sigma^i_A$ is a spacelike
geodesic anchored at its end points.

We would like to use this framework of holographic entanglement entropy for our
case of static bubble geometries. This would allow us to better understand the
holographic interpretation of these bubbles. From the boundary field theory
perspective, dynamics of thin bubble walls in a black hole spacetime is
understood as a thermalisation process~\cite{Balasubramanian:2011ur}. It was
noted in~\cite{Balasubramanian:2011ur}, for the case when $M_{-} = 0$, that
entanglement entropy in these dual field theories shows an interesting 
swallowtail behaviour. This is due to the presence of multiple spatial geodesics that
are anchored at the same boundary interval. We will inspect this behaviour in
detail in the following for arbitrary BTZ masses, focusing on the static
limit. We will see later in \cref{holeography} that this swallowtail behaviour
has important consequences for the holographic reconstruction of bubble
spacetimes.

\subsection{BTZ black holes}

\begin{figure}[t]
\centering
\begin{subfigure}[b]{0.49\textwidth}
\includegraphics[width=\textwidth]{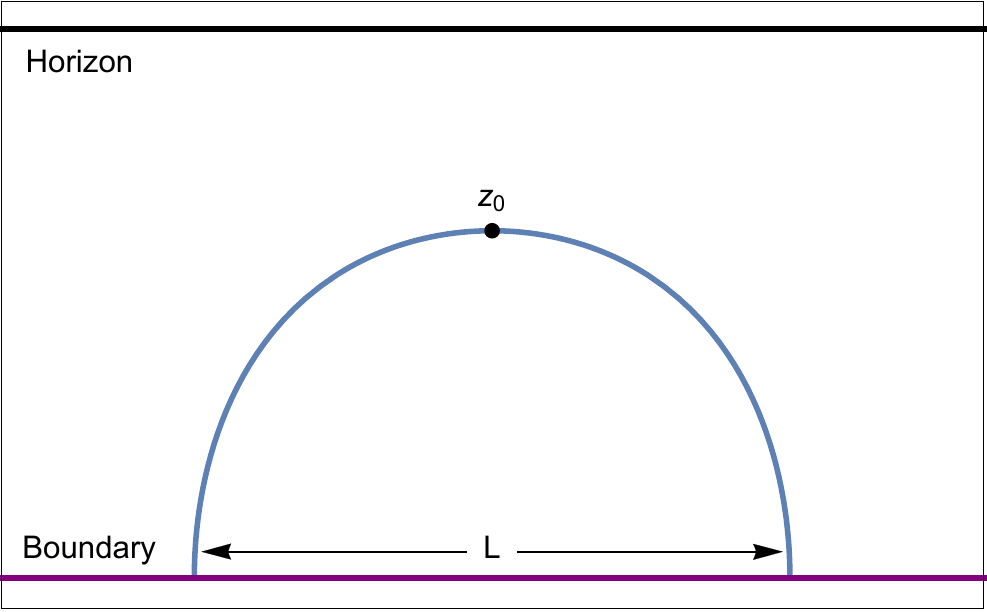}
\caption{}
\end{subfigure}
\begin{subfigure}[b]{0.49\textwidth}
\includegraphics[width=\textwidth]{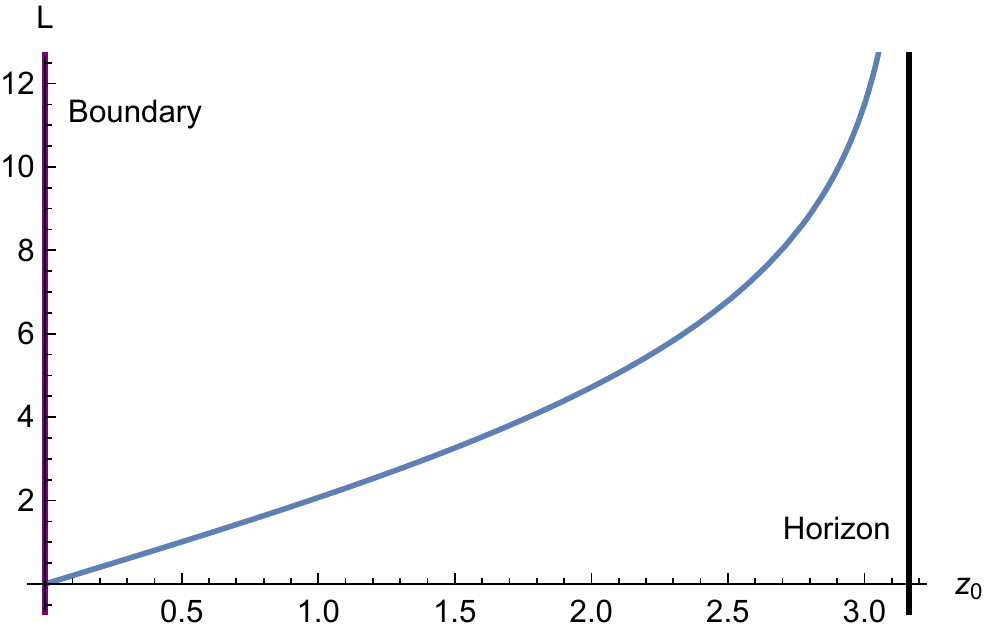}
\caption{}
\end{subfigure}
\caption{\label{BTZ_multi-geodesics} \textbf{(a)} Spatial geodesics anchored
at a boundary interval of length $L$.  \textbf{(b)} Length of the interval
spanned at the boundary $L$ as a function of $z_0$. [Parameters: $\ell = 1$,
$M = 0.1$, and in (a) $t_0 = 0$, $x_0 = 0$, $z_0 = 2$ and $L= 4.71$].}
\end{figure}

\begin{figure}[t]
\centering
\includegraphics[width=0.49\textwidth]{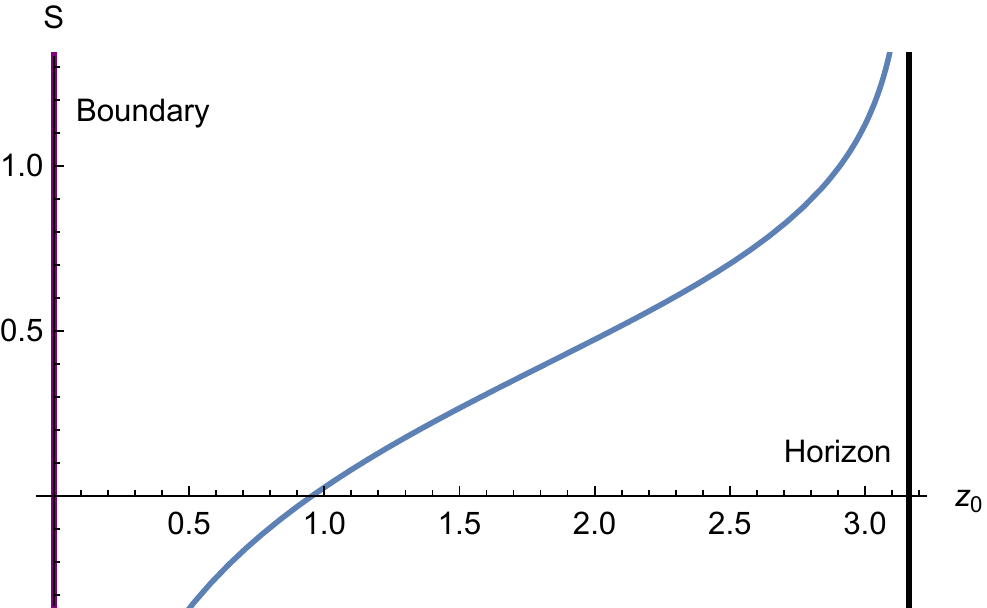}
\includegraphics[width=0.49\textwidth]{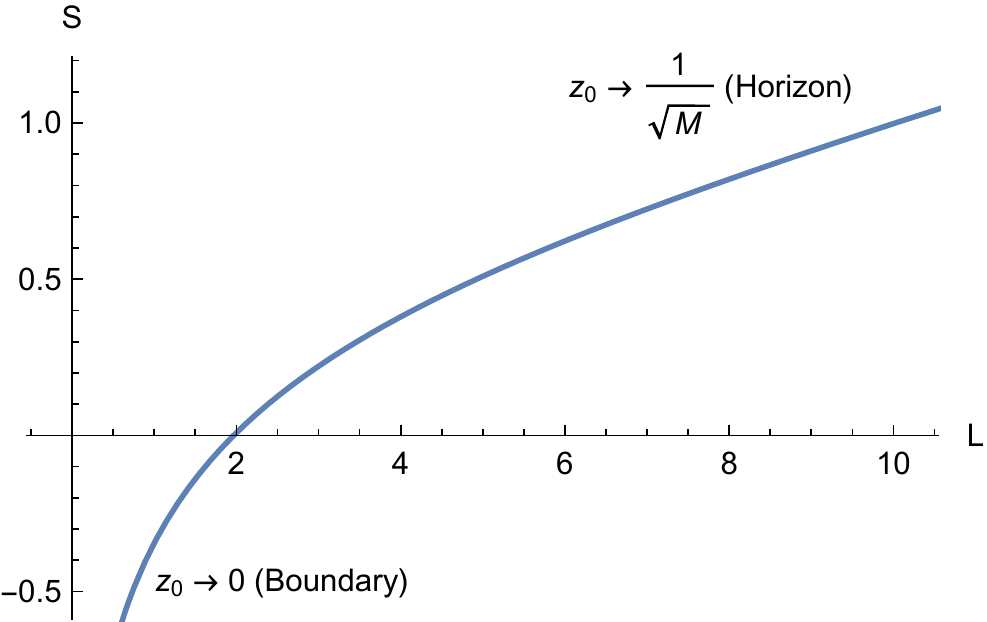}
\caption{\label{BTZentangle} Renormalized holographic entanglement entropy for
BTZ black holes.  The first plot shows the behaviour of renormalized
entangling function $S$ as a function of $z_0$, while the second shows the
mutual behaviour of $S$ with $L$ for varying $z_0$. [Parameters: $\ell = 1$,
$M = 0.1$, $G_3 = 1$].}
\end{figure}

Let us start with a warm-up exercise of computing holographic entanglement
entropy in a bubble-free BTZ black hole spacetime. A detailed analysis can be
found, for example, in~\cite{Ryu:2006bv}. Recall that the metric of a BTZ black
hole is given by \cref{the-metric}. For an interval in the boundary with
endpoints $(t_0,x_0 \pm L/2)$ at the boundary to be linked by a bulk spatial
geodesic, we require
\begin{equation}
L = \frac{2\ell}{\sqrt M}  {\cal{S}}^{-1} \lB 0,z_0,M \rB,
\end{equation}
where the parameter $z_0$ refers to the turning point of the geodesic in the
bulk, which is in defined \cref{geodesics.BTZ}. This constraint has is solved
by\footnote{Mathematically speaking, there is another solution to this
constraint given by
\begin{equation}\nn
z_0 = \frac{1}{\sqrt M}\coth\lb\frac{\sqrt M}{2\ell} L\rb
> \frac{1}{\sqrt{M}}.
\end{equation}
However, this geodesic falls into the black hole and escapes to the other
asymptotic boundary, and hence would not be relevant for us here.
}
\begin{equation}
z_0 = \frac{1}{\sqrt M} \tanh\lb\frac{\sqrt{M}}{~2\ell}L\rb.
\end{equation}
See \cref{BTZ_multi-geodesics} for a diagrammatic representation. We can compute
the length of this geodesics as
\begin{equation}\label{entanglement.BTZ}
S_\e = 2 \ell \log \lb \frac{2 z_0/"e}{\sqrt{1 - M  z_0^2}}\rb + \cO(\e) 
= 2 \ell \log \lb \frac{2/"e}{\sqrt M}\sinh\lb\frac{\sqrt M}{2\ell} L\rb \rb  + \cO(\e),
\end{equation}
where $\epsilon$ is an infinitesimal UV cut-off.  It determines the entanglement
entropy of the interval of length $L$ via the Ryu-Takayanagi formula:
\begin{equation}
S_{\mathrm{EE}}(L) = \frac{c}{3} \log \lb \frac{2\b}{a}\sinh\lb \frac{L}{2\b}\rb\rb,
\end{equation}
where $c = 3\ell/2G$ is the central charge of the CFT, $a=\e\ell$ is the length
scale associated with the UV cutoff in the CFT and $\b = \ell/\sqrt M$ is the
inverse temperature.
See \cref{BTZentangle} for a plot of $S$ verses the turning point $z_{0}$ and
the boundary interval $L$.

\subsection{Static BTZ bubbles}

\begin{figure}[t]
\centering
\includegraphics[width=0.6\textwidth]{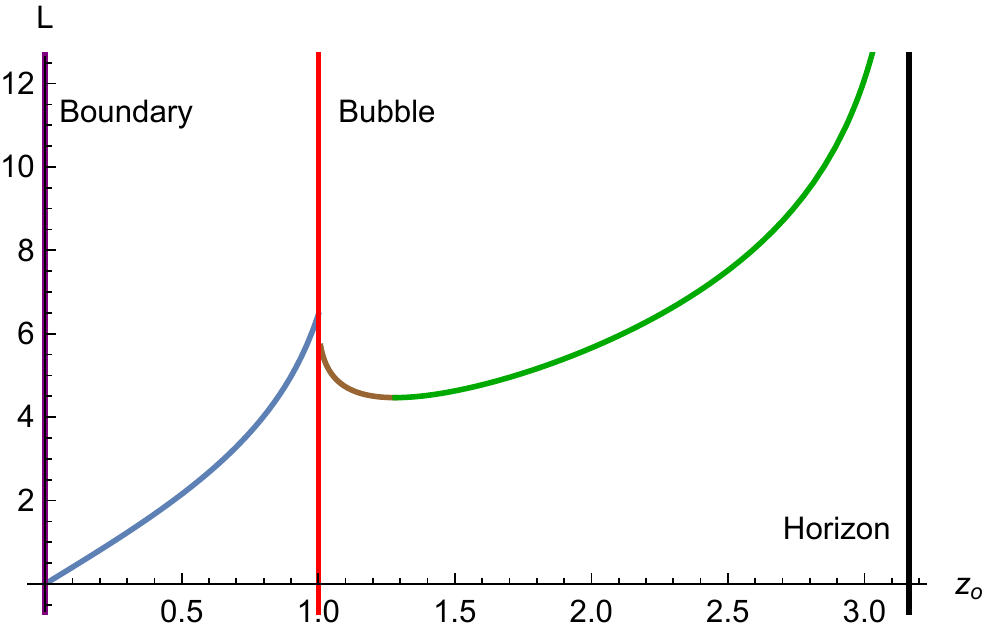} 
\caption{\label{Bubble_multi-geodesic-length} Length of the interval spanned at the
boundary $L$ against the turning point $z_0$. 
[Parameters: $\ell_+ = 2$, $M_+ = 0.8$, $\ell_- = 1$,
$M_- = 0.1$, $Z = 1$].}
\end{figure}

We can now move on to our primary case of interest: static BTZ bubble
spacetimes. \Cref{cross-geodesic-2} implies that for a geodesic to be anchored
at a boundary interval $(t_0,x_0 \pm L/2)$, we must have:
\begin{equation}\label{bigLdef}
L = \frac{2\ell_+}{\sqrt{M_+}} "cS^{-1} \lB 0,z_0,M_+ \rB
+\Theta({\textstyle z_0 - Z}) \lb
\frac{2\ell_-}{\sqrt{M_-}} "cS^{-1} \lB Z,z_0,M_- \rB
- \frac{2\ell_+}{\sqrt{M_+}} "cS^{-1} \lB Z, z_0, M_+ \rB
\rb.
\end{equation}
Unlike the pure BTZ case however, this relation is not simply analytically
invertible, apart from the case when $z_0<Z$. Nevertheless, we can 
qualitatively analyse the behaviour of $L$ as we increase $z_0$.  The first term
in \cref{bigLdef} is an increasing function of $z_0$ until the turning point
$z_0=1/\sqrt{M_+}>Z$ is reached. At that point, the constraints on the
parameters of the plus and minus geometries, \cref{mpmconstraints}, imply that
the second term in \cref{bigLdef} is negative. Specifically, expanding
$z_0=Z + \delta z_{0}$ for small $\delta z_{0}$, we see that
\begin{equation}
L \approx \frac{2\ell_+}{\sqrt{M_+}} "cS[0,Z,M_+] 
+ \frac{2\ell_+ \delta z_{0}}{1-M_+Z^2}
+ \Theta(\delta z_{0}) \sqrt{8Z \delta z_{0}}
\lb \frac{\ell_-}{\sqrt{1-M_- Z^2}} -\frac{\ell_+}{\sqrt{1-M_+ Z^2}} \rb,
\end{equation}
and hence $L(z_0)$ has a local maximum at $z_0=Z$. Meanwhile, as
$z_0\to1/\sqrt{M_-}$, $L\to\infty$, hence $L$ has a minimum between $Z$ and
$1/\sqrt{M_-}$, while for $z_0>1/\sqrt{M_-}$, $L$ decreases again, approaching
zero as $z_0\to\infty$.  This behaviour is displayed in
\cref{Bubble_multi-geodesic-length} for a sample set of parameters for the
bubble geometry.

\begin{figure}[t]
\begin{subfigure}[b]{0.32\textwidth}
\includegraphics[width=\textwidth]{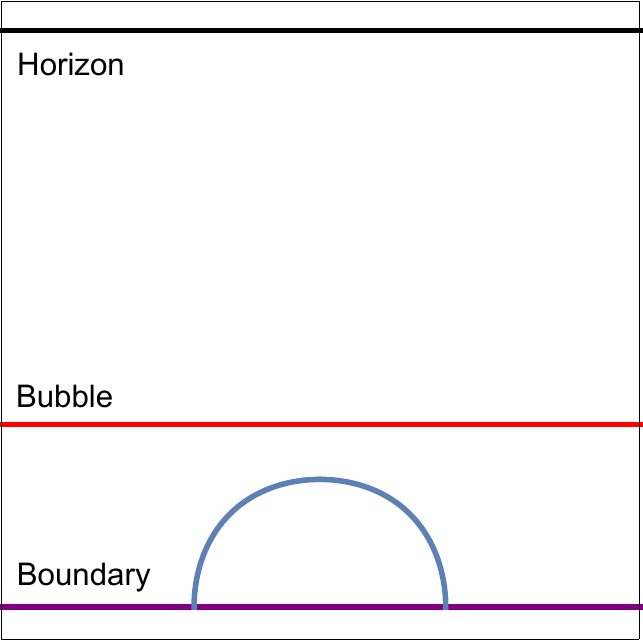}
\caption{}
\end{subfigure}
\begin{subfigure}[b]{0.32\textwidth}
\includegraphics[width=\textwidth]{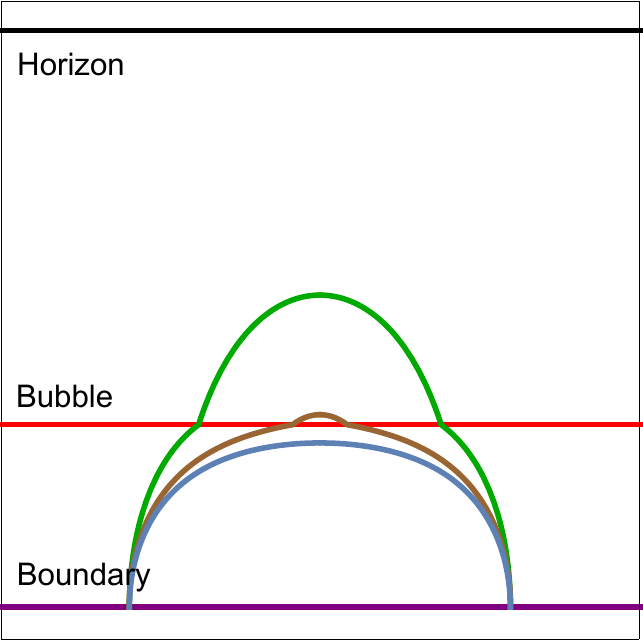}
\caption{}
\end{subfigure}
\begin{subfigure}[b]{0.32\textwidth}
\includegraphics[width=\textwidth]{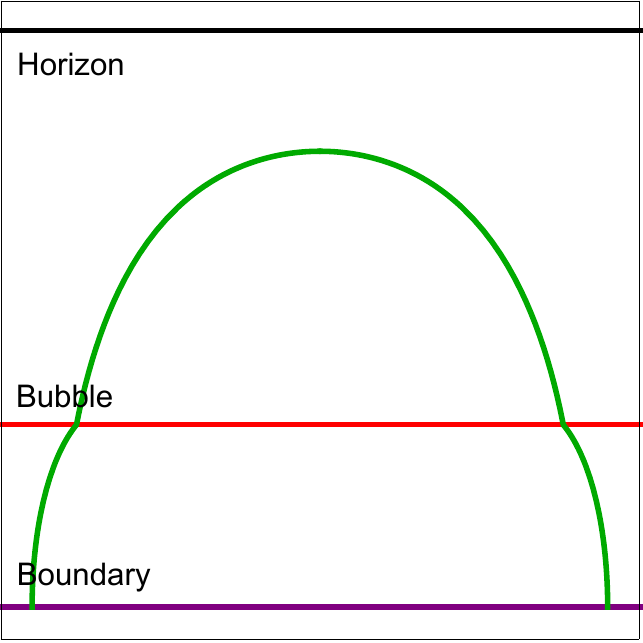}
\caption{}
\end{subfigure}
\caption{\label{Bubble_multi-geodesic} Multiple geodesics anchored at the
boundary.  The geometry parameters are: $\ell_+ = 2$, $M_+ = 0.8$,
$\ell_- = 1$, $M_- = 0.1$, and $Z = 1$. In plot \textbf{(a)} $L=2.15$, with
$z_0 = 0.5$, in \textbf{(b)}, $L=5.07$, with $z_0 = 0.9,~1.05,~1.71$, and in
plot \textbf{(c)}, $L=7.52$, with $z_0 = 2.5$.}
\end{figure}

We can tie this behaviour to the fact that there are now three distinct types of
spatial geodesics, depending on the value of $L$. For small $L$, i.e.
\begin{equation}
L<L_a = \frac{2\ell_+}{\sqrt{M_+}} \tanh^{-1}\lb \sqrt{M_+}Z\rb \, ,
\end{equation}
there are `branch (a)' geodesics, remaining entirely within the ``$+$''
spacetime. As we increase the interval length past some $L = L_{c} < L_{a}$, two
new branches of geodesics pop-up, as it becomes preferable for the geodesic to
cross the bubble wall and take a path through the ``$-$'' spacetime. (see the
middle plot in \cref{Bubble_multi-geodesic}). One of these new geodesics, called
`branch (c)' goes deeper in the bulk than the other, called `branch (b)'. These
geodesics will persist until $L=L_a$, at which point it is no longer possible
for a geodesic to remain in the plus spacetime, and we will simply have a
`branch (c)' geodesic. See the third plot in \cref{Bubble_multi-geodesic}).
\Cref{Bubble_multi-geodesic} shows these different branches of spatial geodesics
as $L$ increases, and \cref{Bubble_multi-geodesic-length} shows a plot of the
length of the geodesics as a function of $z_0$.

\begin{figure}[t]
\centering
\includegraphics[width=0.49\textwidth]{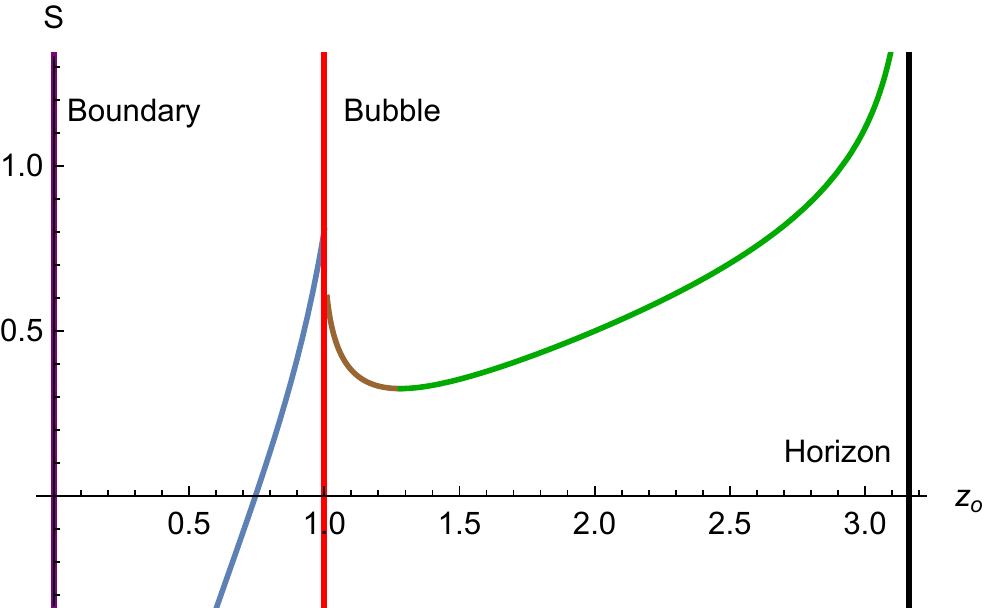}
\includegraphics[width=0.49\textwidth]{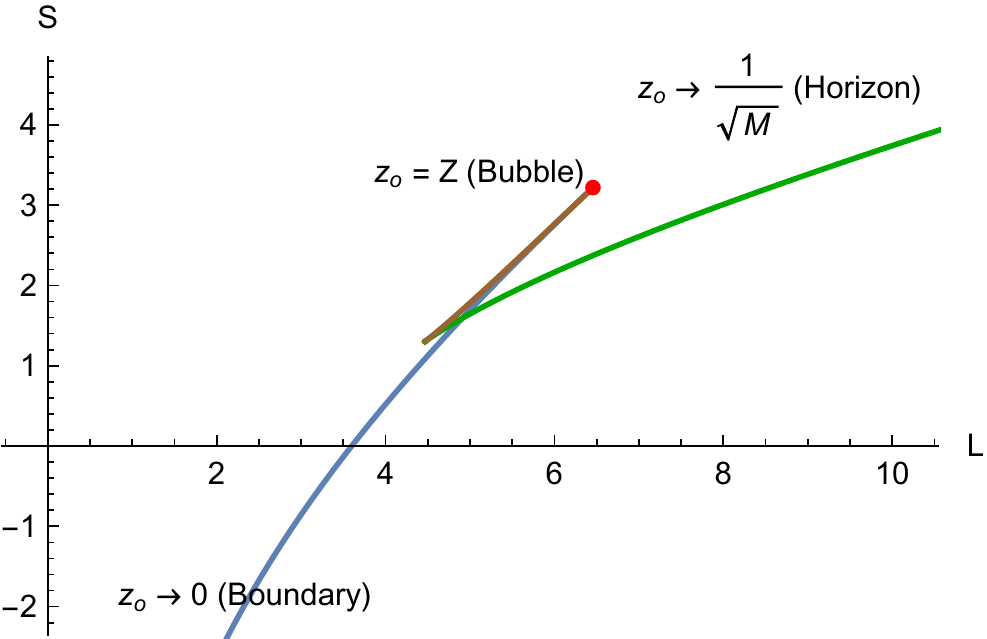}
\caption{\label{BubbleEE} Renormalized holographic entanglement entropy for
BTZ bubbles. The first plot shows the behaviour of the renormalized
entangling function $S$ as a function of $ z_0$, while the second shows the
mutual behaviour of $S$ with $L$ for varying $ z_0$. We see that for a given
$L$ there are 2 to 4 allowed geodesics, the lowest of which computes
entanglement entropy, and the remaining correspond to entwinement. The plot
also shows the swallowtail behavior of holographic entanglement entropy
observed by~\cite{Balasubramanian:2011ur}. [Parameters: $\ell_+ = 2$,
$M_+ = 0.8$, $\ell_- = 1$, $M_- = 0.1$, $Z = 1$, $G_3=1$].}
\end{figure}

The physical appearance of these geodesics suggests an interesting phase
structure as we vary the boundary interval $L$. For small $L$, we expect the
minimal length geodesic to remain `close' to the boundary. However, as $L$
increases and the geodesic gets near to the bubble in the bulk, we might expect
that it becomes preferable for the geodesic to `jump' across the bubble wall,
refracting into the interior ``$-$'' spacetime to cross the bulk at lower
cost. To confirm this suspicion, we need to compare the length of these various
geodesics. In terms of the parameter $z_0$, we find that the length of the
geodesic has the form
\begin{align}\label{Bubble.EntanglementEntropy}
  S_\e
  &= 2\ell_+ \lB \log \lb \frac{2 z_0/"e}{\sqrt{1 - M_+z_0^2}}\rb
    - \Theta( z_0 - Z)\sinh^{-1} \lb\sqrt{ \frac{Z^{-2}-  z_0^{-2}}{ z_{0}^{-2}-M_+}} \rb \rB \nn\\
  &\qquad + 2 \Theta( z_0 - Z) \ell_-
    \sinh^{-1}\lb\sqrt{\frac{Z^{-2}- z_0^{-2}}{ z_{0}^{-2}-M_-}} \rb.
\end{align}
Thus we can readily infer the geometric length of the spacelike geodesics
connecting intervals of length $L$ on the boundary.  In \cref{BubbleEE} we show
the behaviour of the renormalized geodesic length
$S = S_\e + 2\ell_+ \log(\e/2)$ as a function of $z_0$ and $L$.  The geodesic
with least length for a given value of the interval length $L$ determines the
entanglement entropy for that interval, while the longer geodesics can be
interpreted as determining entwinement~\cite{Balasubramanian:2014sra}.

\begin{figure}[t]
\centering
\includegraphics[width=0.5\textwidth]{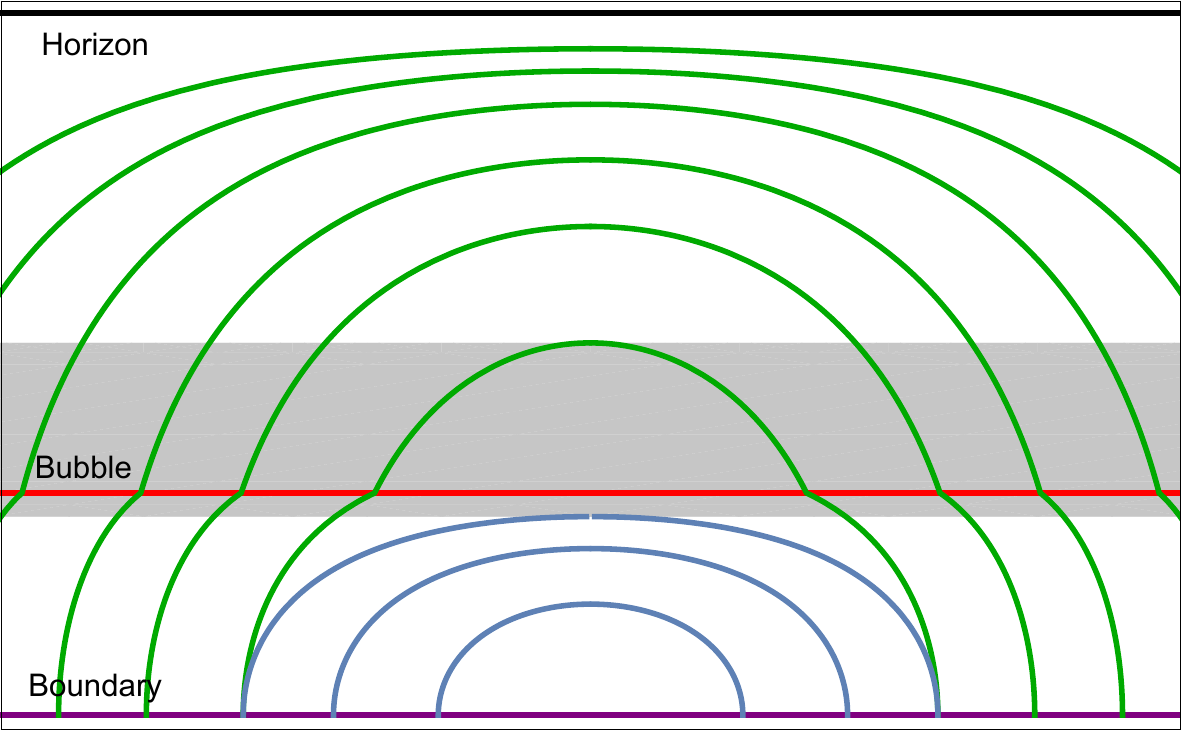}
\caption{\label{bubshad} Two spatial geodesics which are anchored at the same
interval at the boundary. [Parameters: $\ell_+ = 2$, $M_+ = 0.8$,
$\ell_- = 1$, $M_- = 0.1$, $Z = 1$, $t_0 = 0$, $x_0 = 0$, the bubble shadow
is $0.89 < z_0 < 1.68$ and the corresponding critical boundary interval is
$4.91$].}
\end{figure}

The plot in \cref{BubbleEE} clearly shows the phase structure associated to
geodesics in the bubble geometry. As we increase the length of the boundary
interval, entanglement entropy spontaneously jumps from one branch of geodesics
to another. The ``swallowtail'' behaviour of this phase diagram mimics the
Van-der-Waals phase transition in condensed matter systems, and was originally
observed by~\cite{Balasubramanian:2011ur} during the study of collapsing shells
of matter.  As a consequence of this behaviour, there is a range of $z_0$ around
$Z$ for which none of the spatial geodesics are minimal, i.e., all spatial
geodesics with turning points in this region around the bubble correspond purely
to entwinement. We refer to this region as the ``bubble shadow''. Similar
regions, called ``entanglement shadows'', are found around compact BTZ black
holes or a conical deficit~\cite{Czech:2014ppa}. These however are 
fundamentally different from our bubble shadows, as minimal geodesics 
do not penetrate an entanglement
shadow at all, whereas minimal geodesics do in fact cross the bubble shadow
region, they just cannot turn inside it. \Cref{bubshad} illustrates the bubble
shadow effect for our sample bubble geometry.
One might dismiss the occurrence of bubble shadows as an artefact of working
with infinitesimally thin wall, however, one can show that the phenomenon
occurs for smooth walls thinner than a characteristic 
thickness~\cite{PSIStudentsPaper}. 

\section{Holographic reconstruction of static BTZ bubbles}
\label{sec:bubble-reconstruction}

A spacetime manifold is a set of points accompanied with a Lorentzian
metric. The challenge in establishing a holographic dialogue lies in
reconstructing the information that has been projected from the bulk onto the
boundary. In order to reconstruct a bulk spacetime, one requires either
differential information on the bulk, or causal ordering together with a notion
of volume. Here we explore two different methods for reconstructing the bulk,
first a differential approach using a prescription for defining points then a
definition of distance, and second a causal approach, using the structure of
bulk light-cones.

\subsection{Bulk reconstruction using entanglement entropy}\label{holeography}

In~\cite{Balasubramanian:2013lsa,Czech:2014ppa}, a prescription to reconstruct
bulk spacetimes dual to a $1+1$ dimensional CFT was proposed, specifically with
spacetime translation invariance.  Entanglement entropy of an interval in the
boundary CFT is equivalent to the length of the corresponding bulk geodesic
joining the end points. Thus broadly speaking, intervals of different length
sample different depths in the bulk, however a method for extracting the local
spacetime structure is required. The approach of~\cite{Balasubramanian:2013lsa,
Czech:2014ppa} was to first note that a curve $"g("t)$ in the bulk, where $"t$
is an arbitrary parameter on the curve, can be described on the boundary via a
sequence of intervals, each of which is connected by a bulk geodesic that
touches the bulk curve at, and tangent to, a point.\footnote{In general, a
boundary ``function'' $\a(x)$ constructed this way can be multivalued. In that
case, we will need to define it parametrically, i.e. $\a(\zeta)$, $x(\zeta)$
for some parameter $\zeta$. Correspondingly for the differential entropy we
have $E[\a] = \half \int \df\zeta ~ x'(\zeta) S'(\a(\zeta))$.} This sequence
of intervals is expressed as a \emph{boundary function} $\a(x)$, where $x$ is
the centre and $x\pm \a(x)$ are the endpoints of the interval. The bulk curve
$"g("t)$ thus corresponds to a particular function $"a(x)$ on the
boundary. The length of $"g("t)$ has a special interpretation in the bulk; it
computes the so called \emph{differential
entropy}~\cite{Balasubramanian:2013lsa} of $"a(x)$,
\begin{equation}
E["a] \equiv \half \int \df x \left.\frac{\df S(\a)}{\df \a}\right|_{\a = \a(x)} 
=  \frac{\text{Length}["g]}{4G_3}.
\end{equation}
Here $S("a)$ denotes the entanglement entropy (or entwinement) 
corresponding to a boundary interval of length $L = 2"a$.

Having identified functions on the boundary with curves in the bulk, we then
identify a subset of the boundary functions that correspond to bulk points. To
do this, simply note that a point is a limit of a sequence of closed curves with
length and spanning area tending to zero. To be precise, a \emph{point boundary
  function} $"a_p(x)$ corresponding to a point $p$ in the bulk is a sequence of
intervals at the boundary, subtended by the geodesics passing from the point
$p$. Therefore, the point boundary functions are a family of boundary functions
parametrized by 3 parameters --- coordinates of the associated point in the
bulk. By definition, the differential entropy of a point boundary function
vanishes $E["a_p] = 0$, which gives a necessary, but not sufficient,\footnote{To
  be precise, differential entropy is computed by the signed length of the bulk
  curve. Therefore a bulk curve with self intersections might lead to non-zero
  differential entropy if the clockwise length cancels the counter-clockwise
  length of the curve. } condition to determine them.

Point boundary functions are crucial to this bulk-reconstruction scheme, as they
provide a boundary interpretation of the bulk manifold. Furthermore, as showed
by~\cite{Czech:2014ppa}, given two points $p$ and $q$ in the bulk, geodesic
distance between them can be computed using the corresponding point boundary
functions $"a_p(x)$ and $"a_q(x)$
\begin{equation}
  D(p,q) = D["a_p,"a_q] \equiv \half |E[\min("a_p,"a_q)]|.
\end{equation}
The boundary function $\min("a_p,"a_q)(x)$ is defined quite intuitively as
\begin{equation}
  \min("a_p,"a_q)(x) =
  \begin{cases}
    "a_p(x), &\quad "a_p(x) < "a_q(x), \\
    "a_q(x), &\quad "a_p(x) \geq "a_q(x).
  \end{cases}
\end{equation}
Consequently, if we are given the family of point boundary functions, and the
entanglement (plus entwinement) profile of the boundary, we can reconstruct the
bulk spacetime points and their mutual distances, hence recovering the bulk
geometry.

\subsubsection{Point boundary functions for BTZ black holes}

A crucial part of the aforementioned reconstruction scheme is to be able to work
out the family of point-boundary functions for a given CFT, without invoking the
holographic bulk. As we mentioned above, a necessary condition for a boundary
function $"a(x)$ to be a point boundary function is that $E["a] = 0$, however it
is not sufficient. To get some insight into the generic prescription, we start
with the family of point boundary functions when we know that the holographic
dual is a BTZ black hole geometry.

Consider a point $p=(x_p,z_p)$ on a spatial slice of a BTZ black hole with mass
$M$ and AdS radius $\ell$. In \cref{spatial-geodesics} we established that the family
of spatial geodesics can be characterized by their turning point $p_0=(x_0,z_0)$
in the bulk. Requiring these geodesics to pass the point $p$ determines
$|x_0-x_p|$ leaving a family of geodesics depending on $z_0 \geq 0$ and
$\sgn(x_0-x_p)$. By working out the intersection of these geodesics with the
boundary and defining a parameter $\lambda = z_0 \sgn(x_0-x_p)$, point boundary
functions can be expressed parametrically as (see \cref{geodesics.BTZ,bigLdef})
\begin{equation}
\a_p(\lambda)
= \frac{\ell}{\sqrt M} "cS^{-1}[0,\lambda,M], \qquad x(\lambda)
= x_p + \sgn(\lambda) \frac{\ell}{\sqrt M} "cS^{-1}[z_p,\lambda,M].
\end{equation}
On the other hand, the entanglement function for these boundary functions is
given by the length of the geodesics given in \cref{entanglement.BTZ} leading to
\begin{equation}
S(\lambda) = 2 \ell \log \lb \frac{2|\lambda|/\e}{\sqrt{1 - M\lambda^2}}\rb.
\end{equation}

Authors in~\cite{Czech:2014ppa} noted that these point boundary functions obey a
curious relation independent of the black hole parameters $\ell$ and $M$
\begin{equation}\label{K-EOM}
\lb 1 - {\a'(x)}^2\rb \left.\frac{\df^3 S(\a)}{\df \a^3}\right|_{\a= \a(x)}
+ 2 \a''(x) \left.\frac{\df^2 S(\a)}{\df \a^2}\right|_{\a= \a(x)} = 0.
\end{equation}
This is a second order ordinary differential equation for $"a(x)$, and hence
completely determines the two parameter family of point boundary
functions. Given that there is no explicit reference to parameters of the bulk
in this relation, one might wonder if it holds true in general, and if it
could be used as a boundary definition of point boundary functions. 
One can in fact write down an action
\begin{equation}\label{K-action}
I["a] = \int \df x \sqrt{-\lb1 - {\a'(x)}^2\rb \left.
\frac{\df^2 S(\a)}{\df \a^2}\right|_{\a= \a(x)}},
\end{equation}
whose extrema lies at the point boundary functions. From the point of view of a
bulk curve $"g("t)$ associated with $"a(x)$, the action is merely
$\int_{"g}d"t \sqrt{h} K$ where $d"t \sqrt{h}$ is the line element on $"g("t)$,
while $K$ is the extrinsic curvature. Due to the Gauss-Bonnet theorem 
in negatively curved spacetimes, this integral extremizes curves 
enclosing zero area, i.e.\ points.

These hints lead the authors of~\cite{Czech:2014ppa} to conjecture that perhaps
the action \cref{K-action} is more generic and can be used to isolate point
boundary functions in more generic translationally invariant boundary field
theories. In the following subsection, we provide a counter example of this
naive expectation using our BTZ bubbles. In the absence of a universal bulk
independent mechanism to find point boundary functions, this bulk reconstruction
mechanism is incomplete.\footnote{In the language of integral geometries, the
action \cref{K-action} can be thought of as length on an auxiliary space with
metric $ds^2 = \frac{\df^2 S(\a)}{\df \a^2} \lb - dx^2 + d"a^2 \rb$. In a
recent paper on integral geometries~\cite{Czech:2015qta}, the authors mentioned a
different mechanism to work out point boundary functions assuming some strict
conditions on the bulk. Unfortunately, most of these conditions, in particular
the assumption that there are no conjugate points (i.e.\ no two geodesics
intersect at more that one points), break down in the presence of the
bubbles.}

\subsubsection{Point boundary functions for bubbles}\label{sec:pbf_bubbles}

We now turn our attention to point boundary functions for BTZ bubbles, and
inspect if they agree with the naive differential equation (\ref{K-EOM}). Let us
first consider a point $p=(x_p,z_p)$ outside the bubble in ``$+$'' spacetime,
i.e. $z_p < Z$. Geodesics passing through $p$ are of one of the two types:
either they turn outside the bubble staying in ``$+$'' spacetime all the while,
or they penetrate the bubble and turn in ``$-$'' spacetime. Denoting the turning
point by $p_0 = (x_0,z_0)$, we can find the center of the boundary interval
(which is merely $x_0$ by symmetry) using
\cref{cross-geodesic-2,cross-geodesic-1}, and requiring that the geodesic passes
the point $p$
\begin{multline}\label{point-x-ahead-bubble}
x(\lambda) = x_p + \sgn(\lambda) \frac{\ell_+}{\sqrt{M_+}} \bigg(
"cS^{-1} \lB z_p ,\lambda,M_+ \rB
- "Q(|\lambda| - Z) "cS^{-1} \lB Z, \lambda, M_+ \rB \bigg) \\
+ \sgn(\lambda) "Q(|\lambda| - Z) \frac{\ell_-}{\sqrt{M_-}}
"cS^{-1} \lB Z,\lambda,M_- \rB.
\end{multline}
As in the bubble-free case, we have defined $\lambda = z_0 \sgn(x_0-x_p)$.  On
the other hand, for the point $p$ behind or on the bubble in ``$-$'' spacetime,
i.e.  $z_p \geq Z$, every geodesic must turn in ``$-$'' spacetime, hence
$z_0 \geq Z$ as well. In this case, using
\cref{cross-geodesic-2,cross-geodesic-1} the center of the boundary interval is
given as
\begin{equation}\label{point-x-behind-bubble}
x(\lambda) = x_p + \sgn(\lambda) \frac{\ell_-}{\sqrt{M_-}}
"cS^{-1} \lB z_p ,\lambda,M_- \rB.
\end{equation}
Now to specify the point boundary functions, we just need the length of the
boundary interval subtended by a spatial geodesic in terms of the turning
point. We can directly borrow the results from \cref{bigLdef} to get
\begin{multline}\label{point-alpha-bubble}
\a(\lambda)
= \frac{\ell_+}{\sqrt{M_+}} \bigg( "cS^{-1}[0,\lambda,M_+]
- \Q(|\lambda| - Z) "cS^{-1}[Z,\lambda,M_+] \bigg) \\
+ \Q(|\lambda| - Z) \frac{\ell_-}{\sqrt M_-} "cS^{-1}[Z,\lambda,M_-].
\end{multline}
In summary, \cref{point-alpha-bubble} along with \cref{point-x-ahead-bubble} for
a point in front of the bubble ($|\lambda|<Z$) and \cref{point-x-behind-bubble}
for a point behind the bubble ($|\lambda|\geq Z$) parametrically specify the
entire set of point boundary functions for a bubble geometry.

Finally, we compute the entanglement function $S(\lambda)$ by computing 
the length of the aforementioned geodesics. Using 
\cref{Bubble.EntanglementEntropy} we get
\begin{align}\label{entropy-bubble}
S(\lambda) 
&= 2\ell_+ \lB \log\lb \sqrt{\frac{4\lambda^2/\e^2}{1-M_+\lambda^2}} \rb
- \Q(|\lambda| - Z)\sinh^{-1}
\lb\sqrt{ \frac{Z^{-2}- \lambda^{-2}}{ \lambda^{-2}-M_+}} \rb \rB \nn\\
&\qquad + 2 \Q(|\lambda| - Z) \ell_-
\sinh^{-1}\lb\sqrt{\frac{Z^{-2}- \lambda^{-2}}{ \lambda^{-2}-M_-}} \rb.
\end{align}
We wish to see whether the point boundary functions satisfy the bulk independent
differential equations (\ref{K-EOM}). To do so, it is helpful to decompose
\cref{K-EOM} in terms of an intermediate parameter $\lambda$ to give
\begin{equation}\label{K-EOM-parameter}
3 \a''("c) {"a'("c)}^2 + \frac{2}{"c} {"a'("c)}^3
\overset{?}{=}
{x'("c)}^2 \lb \frac{2}{"c} "a'("c)  + "a''("c) \rb + 2 \frac{x''("c)}{x'("c)}
{"a'("c)}^3,
\end{equation}
where all the derivatives are taken with respect to the parameter $"c$.  We have
used the fact that $S'("c) = \frac{2}{|"c|} "a'("c)$ to simplify the equation,
which can be checked to hold explicitly on the point boundary functions
associated with bubble or BTZ geometries. Plugging the point boundary functions
in \cref{point-alpha-bubble,point-x-ahead-bubble,point-x-behind-bubble} into
this differential equation, we can check that it indeed is not satisfied as we
claimed.

Using this counter-example, we see that point boundary functions for generic
holographic CFT's cannot, in fact, be generated using the action in
\cref{K-action}. Another plausible mechanism to figure out point boundary
functions was suggested in the appendix of~\cite{Czech:2015qta}, which among
other things, assumes the bulk to not have any pair of conjugate points (points
connected by more than one geodesics). As we discussed in
\cref{bubble-geodesics}, this assumption severely breaks down in the presence of
bubbles. Bubbles, as we constructed them, exemplify the simplest form of
extended matter in the bulk which confirms with the symmetries of the boundary,
hence we expect that the assumptions in~\cite{Czech:2015qta} will continue to be
invalid in bulk spacetimes with arbitrary matter distribution. Other methods
include using geodesics in boundary Kinematic space as an alternative definition
of point boundary functions, but as we show in \cref{sec:kinematic_space}, they
do not work with bubbles either. We conclude that a more generic
bulk-independent mechanism to define point boundary functions in the field
theory is required for this bulk reconstruction prescription to be complete.

\subsection{Light-cone cuts}\label{LightConeCuts}

In a recent paper~\cite{Engelhardt:2016wgb} (see also~\cite{Engelhardt:2016crc,
Engelhardt:2016vdk}), Engelhardt and Horowitz proposed a new mechanism to
reconstruct the metric of a spacetime, up to a conformal factor, using its
holographic dual. Unlike hole-ography however, which relies on the entanglement
structure of the boundary field theory, this prescription makes use of the
divergence structure of ($d+2$)-point correlation functions in the field theory
to reconstruct the $(d+1)$-dimensional bulk metric. The proposal involves a
novel field theory observable called ``light-cone cuts'' defined as the
hypersurfaces that are null separated from a point in the dual bulk. These cuts
can be used to reconstruct the metric, up to a conformal factor, for a part of
the bulk that is in causal contact with the boundary. From a purely field
theoretic perspective, light-cone cuts can be obtained using the divergence
structure of $(d+2)$-point correlation functions.

In the following, we give a quick review of the reconstruction procedure
of~\cite{Engelhardt:2016wgb}, presented in a slightly different language than
the original material. To a given point $p$ in the causally visible bulk (from
the boundary), we can associate a unique cut $C_p$ at the boundary defined as
the intersection of the boundary with the light-cone of $p$. Engelhardt and
Horowitz further showed that distinct bulk points cannot lead to the same cut,
establishing a bijection between the set of cuts, which we refer to as the
``cut-space'', and points in the causally visible bulk. From the bulk point of
view, it is clear that the cut-space should be a $(d+1)$-parameter family of
hypersurfaces in the $d$-dimensional boundary. Hence a cut in the family can be
represented as $C_{"l}$ where $"l = ("l^0,\ldots,"l^{d})$ is an arbitrary set of
parameters. Given the bijection, $"l$ can also serve as a coordinate system in
the causally visible bulk.  On a given cut $C_{"l}$, we will sometimes choose a
parametrization $"s = ("s^1,\ldots,"s^{d-1})$ and denote a point on the cut as
$C^i_{"l}("s)$ where the index $i$ refers to the boundary coordinates.

One can make a curious observation in the setup defined above. Consider a point
with coordinates $"l_0$ in the causally visible bulk and the corresponding cut
$C_{"l_0}$ at the boundary; if another point $"l$ in the bulk is null separated
from $"l_0$, then the associated cut $C_{"l}$ intersects $C_{"l_0}$ tangentially
at the boundary. That the two cuts intersect follows trivially from the fact
that the light ray joining $"l$ to $"l_0$ hits the boundary at some point $"s_0$
which lies on both the cuts. Furthermore, let $"l$ be in the future of $"l_0$,
then the entire causal future of $"l$ is visible to $"l_0$ while $"l$ can see
the entire past of $"l_0$, and vice-versa if $"l$ is in the past of $"l_0$. If
the cuts were to cross, and not intersect tangentially, at least one of these
conditions will be violated.

The more interesting part of this observation is the converse, which is the
backbone of this reconstruction mechanism. Let $C_{"l_0}$ be a cut at the
boundary, then all the cuts $C_{"l}$ that intersect $C_{"l_0}$ tangentially at
some point $"s_0$, trace out a curve in the $"l$-space (i.e.\ cut-space) which
corresponds to a light ray in the bulk passing through the point $"l_0$ and
hitting the boundary at the point $"s_0$.\footnote{To be precise, the
corresponding curve in the bulk will be a set of two light rays. If
e.g. $C_{"l_0}$ is the future branch of a cut, all the cuts $C_{"l}$ whose
future branches touches $C_{"l_0}$ at some point $"s_0$ will form a light ray
passing through the point $"l_0$ and $"s_0$, while all the cuts whose past
branches touch $C_{"l_0}$ will form another light ray that passes through
$"s_0$ but not $"l_0$. Here however, we will only be interested in the
behaviour of this curve around $"l_0$ and hence will not worry about the
second branch.} It is obviously a curve, as opposed to a higher dimensional
surface, because $"l$ is a set of $(d+1)$ parameters and the condition of
tangential intersection imposes $d$ constraints on it, leaving one free
parameter defining the curve. Since we know that the cuts corresponding to the
points on a light ray joining $"l_0$ with $"s_0$ must be tangential to
$C_{"l_0}$ at $"s_0$, this light ray must be the unknown curve in question.

The philosophy of reconstruction this point forth is rather straightforward: we
assume that we are provided with a family of light-cone cuts $C_{"l}$ at the
boundary with some arbitrary parametrization $"l$. We will come back to the
question of determining this family using the field theory data in a
while. Given a particular cut $C_{"l_0}$ in this family and a point parametrized
by $"s_0$ on $C_{"l_0}$, we will look for a curve $"g_{("l_0,"s_0)}$ in the
$"l$-space which corresponds to cuts tangent to $C_{"l_0}$ at the point
$"s_0$. $"g_{("l_0,"s_0)}$ can be defined via the tangential intersection
constraints
\begin{equation}\label{tangent.equation}
C_{"l}^i("s) = C_{"l_0}^i("s_0), \qquad
\frac{"dd}{"dd"s} C^i_{"l}("s)
\propto \left.\frac{"dd}{"dd"s} C^i_{"l_0}("s)\right|_{"s="s_0} \qquad
\text{for some}~ \s.
\end{equation}
Obviously the point $"l_0$ lies on $"g_{("l_0,"s_0)}$. We denote the tangent
vector to $"g_{("l_0,"s_0)}$ at the point $"l_0$ as $n^a_{("l_0,"s_0)}$, where
the index $a$ runs from $0$ to $d$. We know that from the point of view of the
bulk, $n^a_{("l_0,"s_0)}$ must be a null vector. So we define a metric
$g_{ab}("l)$ on the $"l$-space such that it gives zero norm to
$n^a_{("l_0,"s_0)}$
\begin{equation}\label{cut_metric_constraints}
g_{ab}("l_0) n^a_{("l_0,"s_0)} n^b_{("l_0,"s_0)} = 0.
\end{equation}
We can repeat this procedure for as many values of $"s_0$ as we like, making the
system overdetermined for $g_{ab}("l_0)$. If the boundary field theory is indeed
holographic, there must exist at least one value of $g_{ab}("l_0)$ which
satisfies \cref{cut_metric_constraints} for all values of $"s_0$. Note that
$g_{ab}("l_0)$ has $\half(d+1)(d+2)$ independent components, but a conformal
factor can never be determined through
\cref{cut_metric_constraints}. Nevertheless, we can choose
$\half(d+1)(d+2) - 1 = \half d(d+3)$ generic values of $"s_0$ and determine the
metric $g_{ab}("l_0)$ at the point $"l_0$ up to a conformal factor. We can then
go ahead and repeat this procedure for all values of $"l_0$ to determine the
conformal metric in the entire causally visible bulk.

Now for the reconstruction procedure to be complete, up to a conformal factor,
we just need a field theoretic definition of light-cone
cuts.~\cite{Maldacena:2015iua} argued that an $n$-point Lorentzian correlator in
a holographic field theory can diverge if all its points are null separated from
a bulk point, given that we can associate null momenta to each of the points
while conserving energy-momentum at the bulk point. Let us consider a set of
$d+1$ points $\{x_1,\ldots,x_{d+1}\}$ in a $d$-dimensional holographic field
theory, so that there is a unique point $p$ in the bulk\footnote{In principle,
the point $p$ can also be at the boundary. However, we can easily avoid this
situation by considering time separation between the boundary points large
enough so that they cannot be all null separated from a single point at the
boundary. See~\cite{Engelhardt:2016wgb} for details.}  which is null separated
from all $x_i$. Let us also take two more points $z_1$ and $z_2$ at the
boundary, so that the following correlator diverges,
\begin{equation}
\Big< \mathcal{O}(z_1)\mathcal{O}(z_2)\mathcal{O}(x_1) 
\dots \mathcal{O}(x_{d+1}) \Big> \rightarrow \infty
\label{sing-cor}.
\end{equation}
For this to happen, the point $z_1$ can be anywhere on the light-cone cut
corresponding to the bulk point $p$, while $z_2$ should be another point on the
cut so that the energy-momentum at $p$ is conserved. Now we can find light-cone
cuts by fixing some points $\{x_i\}$ at the boundary --- which fixes the bulk
point $p$ --- and tracing the points $z_1$, $z_2$ at the boundary, so that the
$(d+3)$-point correlator in \cref{sing-cor} remains divergent. Once we have the
cuts, we can go ahead and reconstruct the bulk metric, up to a conformal factor.

Above, we gave an extremely compact review of bulk reconstruction via light-cone
cuts. Naturally, we had to gloss over a lot of tiny yet important details, which
can be found in~\cite{Engelhardt:2016wgb}. In the following we will now try to
explicitly reconstruct our static BTZ bubble using this method. This might help
better understand the prescription by applying to a non-trivial geometry in
presence of matter.

\subsubsection{Light-cone cuts for BTZ black hole}\label{cuts-BTZ}

\begin{figure}[t] 
\centering
\begin{subfigure}[b]{0.44\textwidth}
\centering
\includegraphics[trim={1.5cm 0 0 0},clip,width=0.7\textwidth]{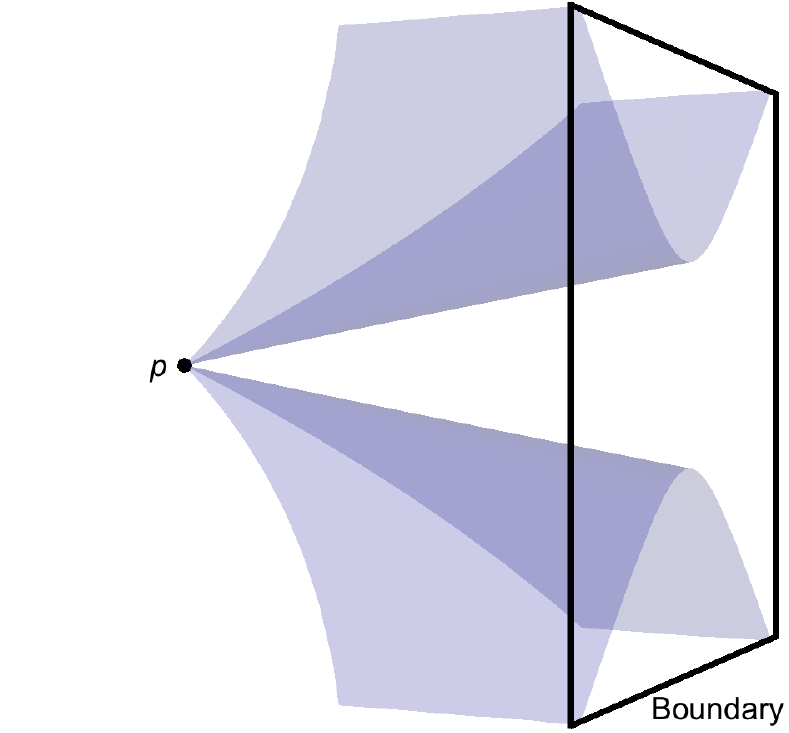}
\caption{}
\end{subfigure}
\hspace{0.04\textwidth}
\begin{subfigure}[b]{0.5\textwidth}
\centering
\includegraphics[width=0.7\textwidth]{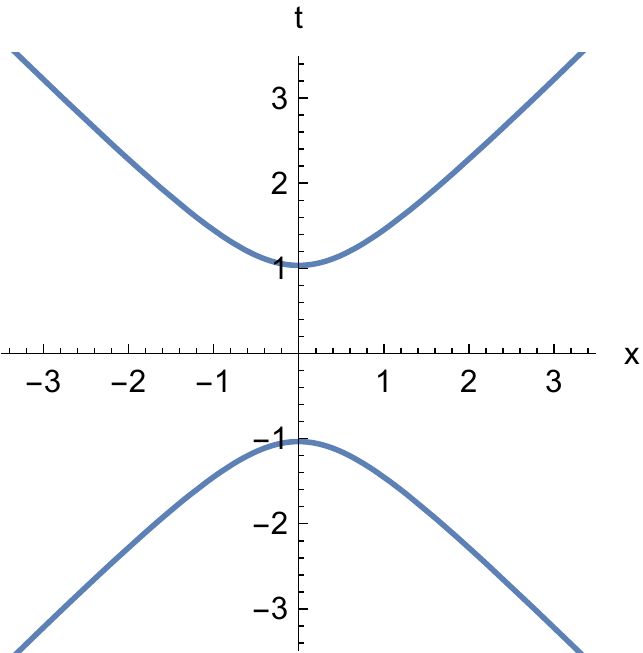}
\caption{}
\end{subfigure}
\caption{\label{fig:BTZ_lc_cuts} \textbf{(a)} Part of the light-cone
originating at a point $p$ in the bulk. \textbf{(b)} A generic light-cone
cut at the boundary. (Parameters: $\ell = 1$, $M = 0.1$ and $p=(0,0,1)$).}
\end{figure}

For simplicity, let us start an ordinary BTZ black hole whose metric has been
given in \cref{the-metric}. A generic null geodesic in this geometry, which
escapes to the boundary, is given in \cref{nullgeod.BTZ}. To find the light-cone
cuts, we need to trace the trajectory of the boundary point $(t_0,x_0)$ in
\cref{nullgeod.BTZ} for all the null geodesics that pass through some point
$(t_p,x_p,z_p)$ in the bulk. We find
\begin{equation}\label{BTZ-cuts}
C^t_{"l}("s) = "l^0 \mp \frac{1}{\sqrt{"m}} "cT^{-1}_t["l^2,"s,"m], \qquad
C^x_{"l}("s) = "l^1 \mp \frac{1}{\sqrt{"m}} "cT^{-1}_x["l^2,"s,"m],
\end{equation}
where $\mu = M/\ell^2$ is the only parameter on which the family of light-cone
cuts depends, and not on the parameters $M$ and $\ell$ independently. This can
be traced back to the fact that under a scaling $M \ra M \O^2$,
$\ell \ra \ell \O$, $ z \ra z/\O$, the metric changes by a conformal factor
$\O^2$, leaving light-cones and hence the light-cone cuts invariant. The set
$"l = ("l^0 = t_p,"l^1 = x_p,"l^2 = \ell z_p)$ can be understood as coordinates
in the cut-space, while the momentum $"s = p$ is a parameter on the
cut. Eliminating $"s$, we can write down a constraining equation for the
light-cone cuts
\begin{equation}
\tanh^2 \lb \sqrt{"m}(C^t - "l^0) \rb
- \lb 1-"m{("l^2)}^2\rb\tanh^2 \lb \sqrt{"m} (C^x - "l^1) \rb
= "m{("l^2)}^2.
\end{equation}
See \cref{fig:BTZ_lc_cuts} for a graphical illustration. Interestingly in the
AdS limit, i.e.\ when the black hole mass $M\ra 0$, light-cone cuts become
hyperbolas,
\begin{equation}\label{pureAdS-cuts}
{(C^t - "l^0)}^2 - {(C^x - "l^1)}^2 = {("l^2)}^2,
\end{equation}
which are independent of the AdS radius $\ell$. This is expected, because in
pure AdS we can always transform away $\ell$ to a conformal factor.

\paragraph*{Reconstruction:} We will now forget about the bulk, and will assume
that these light-cone cuts can be obtained directly by a field theory
computation. Such a computation, in general, might lead to an arbitrarily
different parametrization of the cuts. The metric thus obtained via the
reconstruction prescription, will be related to the one given in
\cref{the-metric} by merely a coordinate transformation and an arbitrary
conformal factor.

Plugging the light-cone cuts in \cref{BTZ-cuts} into \cref{tangent.equation}, a
straightforward computation will lead to the null generators
\begin{equation}
n^a_{("l,"s)} \propto \lb \pm \frac{1}{1- "m {("l^2)}^2}~,~\pm 
"s~,~ \sqrt{1 - "s^2(1-"m {("l^2)}^2)}\rb.
\end{equation}
Defining a metric $g_{ab}\df "l^a \df "l^b$, \cref{cut_metric_constraints} then
takes the form
\begin{multline}
\frac{1}{1- "m {("l^2)}^2} \lb \frac{g_{00}}{1-"m {("l^2)}^2}
+ g_{22} (1-"m {("l^2)}^2) \rb
+ "s^2 \lb g_{11} - g_{22} (1-"m {("l^2)}^2) \rb \\
+ \frac{2 "s g_{01}}{1-"m {("l^2)}^2}
\pm 2 \lb "s g_{12}
+ \frac{g_{02}}{1-"m{("l^2)}^2} \rb \sqrt{1 - "s^2(1-"m {("l^2)}^2)} = 0.
\end{multline}
Since this equation must be imposed for all values of $"s$, we can perform a
Taylor expansion around $"s=0$ and set all the coefficients to zero. For
example, the coefficients of $"s^3$ and $"s^4$ only get contributions from the
last term and set $g_{12} = g_{02} = 0$. The coefficient of $"s$ then sets
$g_{01} = 0$, while the coefficients of $"s^2$ and $1$ determine the metric up
to an arbitrary overall factor
\begin{equation}
\df s^2 = "O^2("l)\lB
- \lb 1 - "m {("l^2)}^2 \rb {(\df"l^0)}^2 + {(\df"l^1)}^2 
+ \frac{1}{1 - "m {("l^2)}^2} {(\df"l^2)}^2 \rB.
\end{equation}
Choosing the conformal factor $"O("l) = \ell/"l^2$ and picking a basis
$\{"l^0 = t,"l^1 = x,"l^2 = \ell z\}$, we can recover the BTZ metric in
\cref{the-metric} with mass $M = "m\ell^2$.

\subsubsection{Light-cone cuts for bubbles}\label{cuts-bubbles}

\begin{figure}[t] 
\centering
\begin{subfigure}[b]{0.255\textwidth}
\centering
\includegraphics[trim={3cm 0 0 0},clip,width=\textwidth]{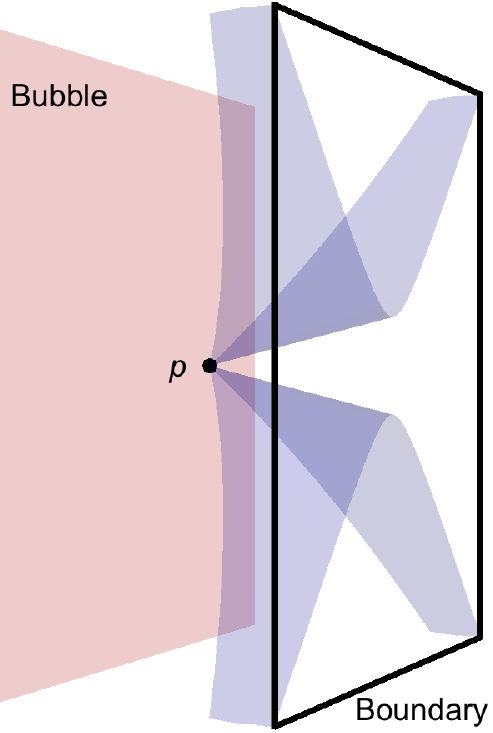}
\caption{}
\end{subfigure}
\begin{subfigure}[b]{0.33\textwidth}
\centering
\includegraphics[trim={1.5cm 0 0 0},clip,width=\textwidth]{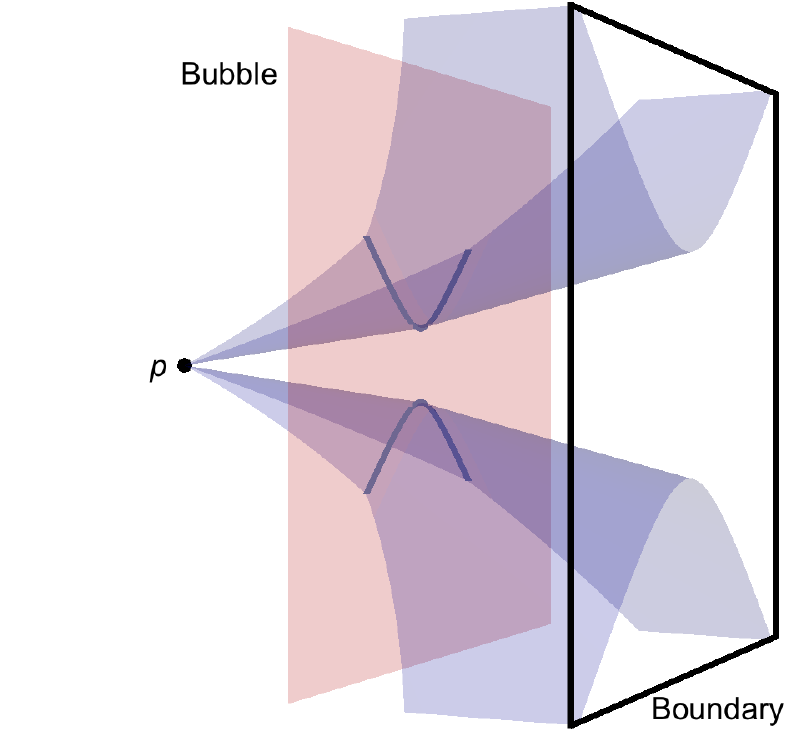}
\caption{}
\end{subfigure}
\hspace{0.02\textwidth}
\begin{subfigure}[b]{0.37\textwidth}
\centering
\includegraphics[width=\textwidth]{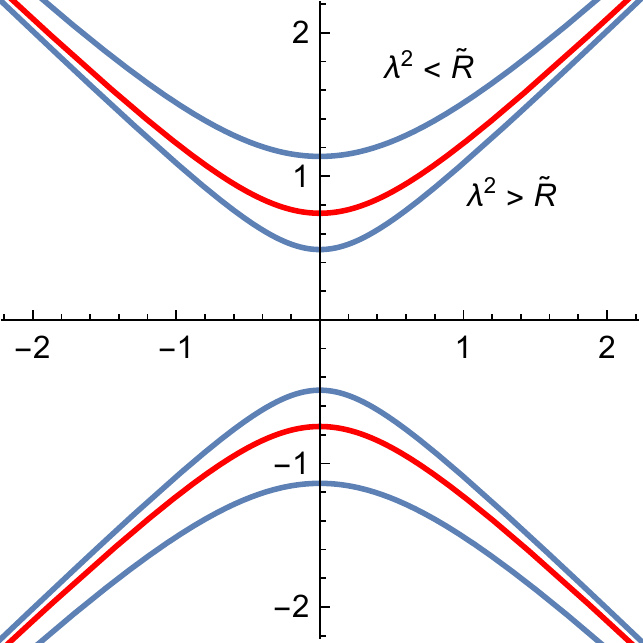}
\caption{}
\end{subfigure}
\caption{\label{fig:Bubble_lc_cuts} Part of the light-cone originating at a
point $p$ in the bulk \textbf{(a)} in front of the bubble and \textbf{(b)}
behind the bubble. \textbf{(c)} A generic set of light-cone cuts at the
boundary; the one in red corresponds to a $"l^2 = R$ (i.e.\ when the
corresponding bulk point is on the bubble). (Parameters: $\ell_+ = 1$,
$M_+ = 0.1$, $\ell_- = 0.5$, $M_- = 0.05$, $R= 0.75$ and
$p=(0,0,0.5), (0,0,1.5)$).}
\end{figure}

We now move on to the case of static BTZ bubbles. Similar to our calculation in
the previous section, we can use null geodesics for bubbles given in
\cref{nullgeod.Bubble.inside,nullgeod.Bubble.outside} to work out the light cone
cuts. We trace the trajectory of the point $(t_0,x_0)$ in
\cref{nullgeod.Bubble.inside,nullgeod.Bubble.outside} while requiring the null
geodesics to pass a fixed point $(t_{p},x_p,z_p)$ in the ``$+$'' part of the
bulk or equivalently $("g t_p,x_p,z_p)$ in the ``$-$'' part. We find
\begin{align}\label{bubble.cuts}
C^{t_+}_{"l}("s)
&= "l^0 \mp \frac{1}{\sqrt{"m_+}} \bigg[
\cT_t^{-1}["l^2,"s,"m_+]
- \Q("l^2-"z_+)\Big(\cT_t^{-1}["l^2,"s,"m_+] - \cT_t^{-1}["z_+,"s,"m_+]
\Big) 
\bigg] \nn\\
&\qquad \mp \Q("l^2 - "z_+) \frac1{"g} \frac{1}{\sqrt{"m_-}}
\Big( \cT_t^{-1}["z"l^2,"g"s,"m_-] - \cT_t^{-1}["z_-,"g"s,"m_-] \Big), \nn\\
C^x_{"l}("s)
&= "l^1 \mp \frac{1}{\sqrt{"m_+}} \bigg[ \cT_x^{-1}["l^2,"s,"m_+]
- \Q("l^2 - "z_+) \lb \cT_x^{-1}["l^2,"s,"m_+] -  \cT_x^{-1}["z_+,"s,"m_+] \rb \bigg] \nn\\
&\qquad \mp \Q("l^2 - "z_+) \frac{1}{\sqrt{"m_-}}
\lb\cT_x^{-1}["z "l^2,"g"s,"m_-] - \cT_x^{-1}["z_-,"g"s,"m_-] \rb.
\end{align}
Here again, we have defined a set of coordinates
$"l = ("l^0 = t_p,"l^1 = x_p,"l^2 = \ell_+ z_p)$ on the cut-space, and
$"s = p_0$ is a parameter on the cut. We have also condensed the parametric
dependence of the cuts into four dimensionless combinations
\begin{equation}
\mu_{\pm} = \frac{M_{\pm}}{\ell_\pm^2}, \qquad
"z_{\pm} = Z \ell_\pm,
\end{equation}
and defined $"z = "z_-/"z_+ = \ell_-/\ell_+$. In terms of these,
$"g = \sqrt{\frac{1-"m_+ "z_+^2}{1-"m_- "z_-^2}}$. The information about one
independent parameter out of $M_\pm$, $\ell_\pm$ and $Z$ is lost. See
\cref{fig:Bubble_lc_cuts}. Interestingly, when the bubble is on an AdS-AdS
interface, i.e. $M_\pm \ra 0$, the light-cone cuts reduce to
\begin{equation}
\begin{split}
\lim_{M_\pm \ra 0}C^{t_+}_{"l}("s)
&= "l^0 \mp \frac{1}{\sqrt{1- "s^2}}\lB "l^2
+ \Q("l^2-"z_+) \lb "z"l^2 - "l^2 - "z_- + "z_+ \rb \rB, \\
\lim_{M_\pm \ra 0} C^x_{"l}("s)
&= "l^1 \mp \frac{"s}{\sqrt{1- "s^2}}\lB "l^2
+ \Q("l^2-"z_+) \lb "z"l^2 - "l^2 - "z_- + "z_+ \rb \rB.
\end{split}
\end{equation}
Note that we can remove all the parametric dependence by performing 
a coordinate transformation in the cut-space:
$"l^2 \to "l'^2 = "l^2 + \Q("l^2-"z_+) \lb "z"l^2 - "l^2 - "z_- + "z_+ \rb$,
after which they merely reduce to their bubble-free hyperbolic form given in
\cref{pureAdS-cuts}. Hence we can infer that the cut-space of an AdS-AdS bubble
is same as that of a bubble-free AdS.

\paragraph*{Reconstruction:} Having obtained the light-cone cuts
\cref{bubble.cuts}, we can now forget about the bulk and try to reconstruct it
using the boundary data. Similar to the ordinary BTZ case, the normal vector
$n^a_{("l,"s)}$ can be obtained by solving \cref{tangent.equation}. We find that
for ``$+$'' spacetime we have
\begin{equation}
n^a_{("l,"s)} \propto \lb ~ \pm \frac{1}{1-"m_+ {("l^2)}^2}~,~
\pm "s ~,~ \sqrt{1 - "s^2\lb 1-"m_+ {("l^2)}^2\rb} ~ \rb.
\end{equation}
while for ``$-$'' spacetime
\begin{equation}
n^a_{("l,"s)} \propto \lb ~ \pm \frac{1/"g}{1-"m_- {("z "l^2)}^2} ~,~
\pm "g"s ~,~
\frac{1}{\z} \sqrt{1 - "g^2"s^2\lb 1-"m_- {("z"l^2)}^2\rb} \rb.
\end{equation}
Putting this back in \cref{cut_metric_constraints} we get the constraints: in
the ``$+$'' spacetime
\begin{multline}
\frac{1}{1- "m_+ {("l^2)}^2}
\lb \frac{g_{00}}{1- "m_+ {("l^2)}^2} + g_{22} \lb 1- "m_+ {("l^2)}^2\rb \rb
+ "s^2\Big( g_{11} - g_{22}\lb 1-"m_+ {("l^2)}^2\rb \Big) \\
\pm 2 \lb
\frac{g_{02}}{1- "m_+ {("l^2)}^2} + "s g_{12} \rb
\sqrt{1 - "s^2 \lb 1-"m_+ {("l^2)}^2\rb }
+ \frac{2"s g_{10}}{1-"m_+{("l^2)}^2} = 0,
\end{multline}
and in the ``$-$'' spacetime
\begin{multline}
\frac{1}{1-"m_- {("z"l^2)}^2} \lb \frac{g_{00}/"g^2}{1-"m_- {("z"l^2)}^2} 
+ g_{22} \frac{1- "m_- {("z"l^2)}^2}{"z^2} \rb
+ "g^2"s^2\lb g_{11} - g_{22}\frac{1-"m_- {("z"l^2)}^2}{"z^2} \rb \\
\pm \frac{2}{"z} \lb \frac{g_{02}/"g}{1-"m_- {("z"l^2)}^2} + "g"s g_{12} \rb
\sqrt{1 - "g^2"s^2\lb 1-"m_-{("z"l^2)}^2\rb}
+ \frac{2 "s g_{10}}{1-"m_- {("z"l^2)}^2} = 0.
\end{multline}
Again performing a Taylor expansion in $"s$ we can find that the metric must be
diagonal. In fact up to a conformal factor $"O^2("l)$ we get
\begin{equation}
ds^2 =
\begin{cases}
\dsp \O^2("l) \lB - \lb 1-"m_+ {("l^2)}^2 \rb {(d"l^0)}^2 +  {(d"l^1)}^2
+ \frac{1}{1-"m_+ {("l^2)}^2} {(d"l^2)}^2 \rB, & "l^2 \geq "z_+ \\
\dsp \O^2("l) \lB - \lb 1-"m_- {("l^2)}^2 \rb {("g d"l^0)}^2
+  {(d"l^1)}^2
+ \frac{"z^2}{1-"m_- {("l^2)}^2} {(d"l^2)}^2 \rB, & "l^2 < "z_+. \\
\end{cases}
\end{equation}
Choosing the coordinates
\begin{equation}
"l^0 = t_+ = t_-/"g, \qquad
"l^1 = x, \qquad
"l^2 = \ell_+ z,
\end{equation}
and a conformal factor $"O("l) = \ell_+/"l^2 = 1/z$, we can recover the metric
in \cref{bubble.metric} with AdS radii $\ell_+$ and $\ell_- = "z \ell_+$, masses
$M_\pm = "m_\pm\ell_\pm^2$ and the bubble radius
$Z = "z_+/\ell_+ = "z_-/\ell_-$.

\section{Discussion}\label{sec:discussion}

In this paper we considered the question of holographic reconstruction of
spacetimes containing non-trivial matter. We believe that this is an important
consistency check for any bulk reconstruction prescription, which aims to build
non-trivial holographic spacetimes using purely boundary observables. Realising
that most of the previous work on geometric bulk reconstruction has concerned
itself with spacetimes which are quotients of AdS, we chose to work with thin
BTZ bubble walls as an example of non-trivial matter content in the bulk. Such a
choice of the matter presence has several unique features. On the one hand, it
has almost the same amount of symmetries as pure AdS or BTZ black hole, whereas
the presence of matter is non-local and the geometry is no longer merely a
quotient of AdS. To retain some analytic control over the problem, we have
further restricted ourselves to static thin bubble wall solutions. Some readers
might argue against such a matter presence as we are explicitly giving away
smoothness of the spacetime manifold. However, thin wall bubbles are a well
known approximation to the more realistic thick wall solutions, for which the
manifold is smooth everywhere. One can easily imagine a sequence of limiting and
still smooth thick wall solutions approximating a thin bubble wall. We therefore
expect the qualitative features discussed in this paper to be valid for thick
bubble walls which are sufficiently thin compared to relevant length scales in
the problem. We intend to discuss this issue in detail in a follow-up
work~\cite{PSIStudentsPaper}.

An additional advantage of working with bubble spacetimes is that they can
provide a toy model for understanding the process of matter collapse and black
hole formation in the bulk. From the boundary perspective, this process is
understood as thermalisation of a field theory state towards its thermal
equilibrium~\cite{Danielsson:1999fa, Balasubramanian:2011ur}. On the one hand,
looking at these non-trivial dynamical processes from the perspective of
holographic bulk reconstruction is expected to bring some new and important
insights in the long-standing puzzles in (quantum) black hole
physics~\cite{Almheiri:2012rt}. On the other hand, they should also guide our
efforts towards writing with a universally applicable bulk reconstruction
scheme. In any case, it would be interesting to extend the analysis done in this
paper to include dynamical thin bubble walls and explore if we find any
qualitative differences in the results. Having sacrificed the time translation
invariance of our setup, it is likely that explicit results will require
numerical techniques. We plan to return to this question in the near future.

We considered two recent schemes of geometric bulk reconstruction. 
Using the hole-ography method, we are able to reconstruct the metric on 
a spatial slice of the bulk using the entanglement structure of the boundary, 
provided we are given a bulk-independent mechanism to work out 
point boundary functions for a given field theory. Unfortunately, as we 
discussed in the bulk of this paper, the current mechanism in place to 
work out these functions using a variational principle seems to break 
down when applied to bubble geometries. There have also been some 
suggestions (see e.g.~\cite{Czech:2016xec}) to use boundary Kinematic
spaces to work out the point boundary functions. However, we show in
\cref{sec:kinematic_space} that this also does not work with bubble 
geometries. In absence of such a mechanism, the hole-ographic 
prescription by itself is incomplete, which adds to the limitations of 
hole-ography previously pointed
out in~\cite{Engelhardt:2015dta}.

The light-cone cuts method of~\cite{Engelhardt:2016wgb}, on the other 
hand, seems to work quite well for bubble spacetimes, considering the 
manifold we are working with is not smooth due to the presence of a 
thin wall. It should be noted, as the authors pointed out themselves, 
that this bulk reconstruction prescription only returns the metric up to 
a conformal factor. This essentially means that the information about 
the volume measure is not recoverable in this scheme. In particular, 
this implies that the light-cone cuts method is ignorant of a thin bubble 
wall bubble between two empty AdS spacetimes, which is conformally
related to an empty AdS. We know that from the boundary field theory
perspective, the presence of a dynamical thin shell in an empty AdS spacetime
corresponds to an RG flow in the boundary field theory, whereas an empty AdS
spacetime corresponds to the vacuum state. Therefore, a lot of such interesting
physics is lost in the light-cone cut prescription, unless we can complement it
with an independent prescription to compute the volume measure. As a future
direction, it would be interesting to explore if the two methods of bulk
reconstruction considered here can be made to complement each other, so as 
to mutually overcome their shortcomings.

Another important direction in the bulk reconstruction program is the ongoing
research on tensor networks~\cite{Almheiri:2014lwa, Pastawski:2015qua,
  Czech:2015kbp}. These methods, again, have been quite successful in describing
the emergence of locally AdS geometries from boundary field theory data. The
natural next step therefore, is to extend this discussion to the cases where
non-trivial matter is present in the bulk. A viable toy model to explore this
direction is provided by the thin bubble walls discussed in this paper, whose
analysis has already been initiated in~\cite{Czech:2016nxc}. Another bulk
restriction prescription that we have not considered in this work is using the
quantum error correcting structure of AdS/CFT, proposed recently
by~\cite{Sanches:2017xhn}. Like the light-cone prescription, it promises to be
able to reconstruct the bubble spacetime up to a conformal factor.

During our holographic analysis of thin bubble walls in \cref{sec-HEE}, we
observed the existence of the so called ``bubble shadows'': a region around a
bubble wall in the bulk spacetime which is only partially probed by minimal
geodesics. These appear to be a generalisation of entanglement shadows found
around BTZ black holes and conical defects, which are spacetime regions where no
minimal geodesics can enter. These preliminary results seem to suggest that such
shadows in boundary entanglement structure might be a generic feature of the
presence of matter in the bulk. However, more analysis is required before these
suggestions can be turned into concrete claims.

\acknowledgements

We would like to thank Mike Appels, Leo Cuspinera, Aurora Ireland 
and Simon Ross for helpful discussions. The work of PB was supported by 
the ``Quantum Universe'' I-CORE
program of the Israel Planning and Budgeting Committee. RG is supported in part
by STFC Consolidated Grant ST/P000371/1. 
AJ is supported by Durham Doctoral Scholarship from Durham University.
AJ would also like to acknowledge the hospitality provided by Perimeter 
Institute for Theoretical Physics, Canada
where part of this project was done. 
Research at Perimeter Institute is supported by the Government of
Canada through the Department of Innovation, Science and Economic 
Development Canada and by the Province of Ontario through the
Ministry of Research, Innovation and Science.

\appendix

\section{Kinematic space for BTZ bubbles}
\label{sec:kinematic_space}

In this appendix we discuss the kinematic spaces associated with 
the bubble spacetimes discussed in this paper. For a detailed discussion 
of Kinematic spaces and their relevance in bulk reconstruction, 
see~\cite{Czech:2015qta, Bhowmick:2017egz, Czech:2016xec} and references
therein. A kinematic space, from the boundary field theory perspective, is
defined as the space of pair of boundary points. In pure AdS, it can
equivalently be defined from the bulk perspective as the space of bulk geodesics
anchored at those boundary points. In more complicated spacetimes however, this
equivalence runs into some trouble because of the existence of multiple
geodesics corresponding to same boundary intervals. For example, AdS spacetime
with a conical defect admits multiple geodesics anchored at the same boundary
points, labelled by their winding number around the defect. The kinematic space
for this geometry was studied in~\cite{Cresswell:2017mbk}. A similar story holds
true for cyclically identified BTZ black holes as well, wherein the geodesics
wind around the horizon instead of the defect.

As we have explored in this paper, BTZ black holes with bubble walls admit
additional geodesics for a subset of boundary intervals, which in a sense are
more non-trivial than the ones wrapping around the horizon. To isolate this
effect, we specialised to planar BTZ black holes, so that we can concentrate on
only the multiple geodesics arising due to the bubble. In this section we would
like to explore Kinematic spaces for these bubbles. Let us start with a generic
discussion of $(2+1)$-dimensional bulk spacetimes, whose constant time slices
look like
\begin{equation}\label{generic-spatial-metric}
"dd s^2 = \frac{1}{z^2} \Big("dd x^2 + f^2(z) "dd z^2 \Big),
\end{equation}
where $f(z) = f(-z)$. In the case of BTZ bubbles discussed in the bulk of this
paper (see metric \bref{bubble.metric}), $f(z)$ takes a step function profile
\begin{equation}
f^{2}(z) = \frac{\ell^{2}_{+}}{1-M_{+}z^{2}}\Theta(Z-z)
+ \frac{\ell^{2}_{-}}{1-M_{-}z^{2}}\Theta(z-Z).
\end{equation}
Spatial geodesics corresponding to metric \bref{generic-spatial-metric} are
given by a two parameter family
\begin{equation}\label{orientedGeodesics}
x("z) = x_0
+ \sgn("z)\sgn(z_0) \int_{|z_0| - |"z|}^{|z_0|}
"dd "l \frac{"l f("l)}{\sqrt{z_0^2 - "l^2}},
\qquad
z("z) = |z_0| - |"z|.
\end{equation}
where $x_0,z_0\in"bbR$. Note that the way we have parametrised this family of
geodesics, it is left invariant by $z_0\to-z_0$ provided we take the parameter
on the geodesic $"z \to -"z$. Therefore every geodesic is counted
twice.\footnote{Eliminating $"z$ and assuming $z_0\geq 0$, these geodesics could
also be written as
\begin{equation}
x(z) = x_0 \pm \int_{z}^{z_0} "dd "l \frac{"l f("l)}{\sqrt{z_0^2 - "l^2}},
\end{equation}
but one would need to take care of two branches, as there are two values of
$x$ for every value of $z < z_0$.} In this sense, \cref{orientedGeodesics}
actually parametrise the set of ``oriented spatial geodesics'', where the
orientation is defined by $\sgn\lb x'("z) \rb = \sgn(z_0)$. This set is
generally known as the ``Kinematic space''. The pair of parameters $(x_0,z_0)$
serve as a basis on the Kinematic space. Locally, we can also use as basis the
$x$-coordinates $(u,v)$ of the points at which the geodesic hits the boundary
$z\to 0$, i.e.  $"z = \pm |z_0|$. They are given in terms of $(x_{0},z_{0})$ as
\begin{align}
u &= x(-|z_0|) = x_0
- \sgn(z_0) \int_{0}^{|z_0|} "dd "l \frac{"l f("l)}{\sqrt{z_0^2 - "l^2}}, \nn\\
v &= x(|z_0|) = x_0
+ \sgn(z_0) \int_{0}^{|z_0|} "dd "l \frac{"l f("l)}{\sqrt{z_0^2 -"l^2}}.
\end{align}
Alternatively, we could also use the local basis $("a,x_0)$ where
\begin{equation}
"a(z_0) = \frac{v-u}{2} = \sgn(z_0)
\int_{0}^{|z_0|} "dd "l \frac{"l f("l)}{\sqrt{z_0^2 - "l^2}},
\end{equation}
is half the signed length of the interval spanned by the geodesic. Note however,
that when there are multiple spatial geodesics corresponding to the same
boundary interval, such bases are not globally well defined. One can define a
measure on the Kinematic space locally via the Crofton form\footnote{The measure
is defined via the requirement that the length of a closed curve $"g$ in the
bulk can be reproduced by a Kinematic space integral
\begin{equation}
\frac{1}{4G_N}\text{Length}["g] = \frac{1}{4} \int
"o_{\text{KS}} n_{"g},
\end{equation}
where $n_{"g}$ is the number of times a given geodesic intersects the curve
$"g$~\cite{Balasubramanian:2013lsa}.}~\cite{Czech:2015qta}
\begin{equation}
"o_{\text{KS}}
= \frac{\dow^2 S(u,v)}{\dow u \dow v} "dd u \wedge "dd v
= -\frac12 \frac{"dd^2 S("a)}{"dd "a^2} "dd x_0 \wedge "dd "a,
\end{equation}
where $S(u,v) = S("a)$ is the length of the geodesic being considered. In
accordance with the symmetries of our setup, we have taken $S("a)$ to be only
dependent on the length of the boundary interval and not its location. In terms
of the global coordinates $(x_0,z_0)$, the Crofton form is given as
\begin{equation}
"o_{\text{KS}} = -\frac12 \frac{"dd}{"dd z_0}\bfrac{S'(z_0)}{"a'(z_0)}
"dd x_0 \wedge "dd z_0.
\end{equation}
where $S(z_0)$ can be computed to be
\begin{equation}
S(z_0) = 2\sgn(z_0)\int_0^{|z_0|}"dd"l \frac{|z_0|f("l)}{"l\sqrt{z_0^2 - "l^2}}.
\end{equation}
The authors in~\cite{Czech:2015qta} further endowed the Kinematic space with a
causal structure via the metric represented locally as
\begin{equation}
"dd s^2_{\text{KS}} = 2\frac{\dow^2 S(u,v)}{\dow u \dow v} "dd u "dd v
= -\frac12 \frac{\dow^2 S(\alpha)}{\dow\alpha^{2}} \lb - "dd "a^2 + "dd x_0^2 \rb.
\end{equation}
In our $(x_0,z_0)$ coordinate system, the same turns into
\begin{equation}\label{bubKinSpace}
"dd s^2_{\text{KS}}
= -\frac12 \frac{"dd}{"dd z_0}\bfrac{S'(z_0)}{"a'(z_0)} \lb - "a'(z_0)"dd
z_0^2 + \frac{1}{"a'(z_0)} "dd x_0^2 \rb.
\end{equation}
We would like to inspect this metric on the Kinematic space for our bubble
setup. $\alpha(z_{0})$ and $S(z_{0})$ for these bulk geometries have been given
in \cref{point-alpha-bubble} and \cref{entropy-bubble} respectively. Taking a
straightforward derivative we can find that
\begin{gather}
-\frac12 \frac{"dd}{"dd z_0}\bfrac{S'(z_0)}{"a'(z_0)} = \frac{1}{z_0^2}, \\
"a'(z_0) = \frac{\ell_+}{1-M_+ z_0^2} + \frac{z_0 "Q(z_0 - Z)}{\sqrt{z_0^2 - Z^2}}
\lb \frac{\ell_- \sqrt{1 - M_- Z^2}}{1-M_- z_0^2}
- \frac{\ell_+ \sqrt{1 - M_+ Z^2}}{1-M_+ z_0^2}\rb.
\end{gather}
The first thing we immediately note is that $\alpha'(z_{0})$ is not well defined on
the bubble wall $z_{0} = Z$. But that is hardly surprising, as we are working
with a thin bubble wall. We expect this singularity to go away when we work with
a smooth wall instead. However, $\alpha'(z_{0})$ also vanishes at some point
$z_{0}>Z$ finite distance away from the bubble, which we cannot attribute to
working with a thin wall. It is not just a coordinate singularity either, scalar
curvature $R$ blows up at this point, indicating that there is something really
wrong with the spacetime. In fact, inspecting the behaviour or $R$ as a function
of $z_{0}$, we see that the geometry in question is not nice at all.

One of the motivations of working with Kinematic spaces in the context of
holographic bulk reconstruction is that the geodesics in Kinematic space are
conjectured to correspond to point boundary functions in boundary field theory
(see \cref{holeography} for the definition of point boundary functions). If
true, this could provide the missing piece in the puzzle for hole-ographic bulk
reconstruction. However, the geodesic equation for the Kinematic space metric
\bref{bubKinSpace} is given in \cref{K-EOM}, and as we discussed in
\cref{sec:pbf_bubbles}, is not satisfied for point boundary functions in bubble
spacetimes. 


\bibresources{mySpires_ajain,Reconstruction_rev}

\makereferences

\end{document}